\documentclass[twoside,11pt]{article}

\usepackage{jmlr2e}

\usepackage{amsmath}
\usepackage{amssymb}
\usepackage{amsthm}
\usepackage{graphicx}
\usepackage{color}
\usepackage{mathrsfs}

\usepackage{epsfig}
\usepackage{natbib}
\usepackage[linesnumbered,ruled]{algorithm2e}

\makeatletter
\def\widebreve{\mathpalette\wide@breve}
\def\wide@breve#1#2{\sbox\z@{$#1#2$}%
     \mathop{\vbox{\m@th\ialign{##\crcr
\kern0.08em\brevefill#1{0.8\wd\z@}\crcr\noalign{\nointerlineskip}%
                    $\hss#1#2\hss$\crcr}}}\limits}
\def\brevefill#1#2{$\m@th\sbox\tw@{$#1($}%
  \hss\resizebox{#2}{\wd\tw@}{\rotatebox[origin=c]{90}{\upshape(}}\hss$}
\makeatletter

\newcommand{\be}{\begin{equation}}
\newcommand{\ee}{\end{equation}}
\newcommand{\bes}{\begin{equation*}}
\newcommand{\ees}{\end{equation*}}
\newcommand{\beqn}{\begin{eqnarray}}
\newcommand{\eeqn}{\end{eqnarray}}
\newcommand{\beqns}{\begin{eqnarray*}}
\newcommand{\eeqns}{\end{eqnarray*}}
\newcommand{\lkr}{\left(}
\newcommand{\lkv}{\left[}
\newcommand{\rkv}{\right]}
\newcommand{\rkr}{\right)}
\newcommand{\lfi}{\left\{}
\newcommand{\rfi}{\right\}}

\newcommand{\Del}{\Delta}

\newcommand{\eps}{\epsilon}

\newcommand{\ga}{\gamma}

\newcommand{\lam}{\lambda}

\newcommand{\sig}{\sigma}

\newcommand{\Om}{\Omega}

\newcommand{\Te}{\Theta}

\newcommand{\EE}{\ensuremath{{\mathbb E}}}

\newcommand{\PP}{\ensuremath{{\mathbb P}}}

\newcommand{\RR}{{\mathbb R}}

\newcommand{\vect}{\mbox{vec}}

\newcommand{\diag}{{\rm diag}}

\newcommand{\Tr}{\mbox{Tr}}

\newcommand{\rank}{\mbox{rank}}

\newtheorem{thm}{Theorem}
\newtheorem{lem}{Lemma}
\newtheorem{cor}{Corollary}
\newtheorem{rem}{Remark}

\newcommand{\bB}{\boldsymbol{B}}

\newcommand{\bbX}{\boldsymbol{X}}

\newcommand{\bd}{\mathbf{d}}

\newcommand{\bq}{\mathbf{q}}

\newcommand{\bA}{\mathbf{A}}
\newcommand{\boB}{\mathbf{B}}
\newcommand{\bC}{\mathbf{C}}
\newcommand{\bD}{\mathbf{D}}

\newcommand{\bF}{\mathbf{F}}
\newcommand{\bG}{\mathbf{G}}
\newcommand{\bH}{\mathbf{H}}
\newcommand{\bI}{\mathbf{I}}

\newcommand{\bO}{\mathbf{O}}
\newcommand{\bP}{\mathbf{P}}
\newcommand{\bQ}{\mathbf{Q}}
\newcommand{\bR}{\mathbf{R}}
\newcommand{\bS}{\mathbf{S}}
\newcommand{\bU}{\mathbf{U}}
\newcommand{\bV}{\mathbf{V}}
\newcommand{\bW}{\mathbf{W}}
\newcommand{\bX}{\mathbf{X}}
\newcommand{\bY}{\mathbf{Y}}
\newcommand{\bZ}{\mathbf{Z}}

\newcommand{\bzero}{\mathbf{0}}
\newcommand{\bone}{\mathbf{1}}

\newcommand{\bPsi}{\mbox{\mathversion{bold}$\Psi$}}

\newcommand{\bLam}{\mbox{\mathversion{bold}$\Lambda$}}

\newcommand{\bTe}{\mbox{\mathversion{bold}$\Theta$}}

\newcommand{\calA}{{\mathcal{A}}}
\newcommand{\calB}{{\mathcal{B}}}

\newcommand{\calF}{{\mathcal{F}}}
\newcommand{\calG}{{\mathcal G}}
\newcommand{\calH}{{\mathcal H}}

\newcommand{\calM}{{\mathcal M}}

\newcommand{\calO}{{\mathcal O}}
\newcommand{\calP}{{\mathcal{P}}}
\newcommand{\calR}{{\mathcal{R}}}

\newcommand{\calW}{{\mathcal W}}
\newcommand{\calX}{{\mathcal{X}}}
\newcommand{\calY}{{\mathcal{Y}}}

\newcommand{\calV}{{\mathcal{V}}} 
\newcommand{\calU}{{\mathcal{U}}}

\newcommand{\sumlL}{\sum_{l=1}^L}
\newcommand{\summM}{\sum_{m=1}^M}

\newcommand{\hbC}{\widehat{\bC}}
\newcommand{\hbG}{\widehat{\bG}}
\newcommand{\hbH}{\widehat{\bH}}
\newcommand{\hbTe}{\widehat{\bTe}}
\newcommand{\hbD}{\widehat{\bD}}
\newcommand{\hbU}{\widehat{\bU}}
\newcommand{\hbV}{\widehat{\bV}}
\newcommand{\hbW}{\widehat{\bW}}
\newcommand{\hbQ}{\widehat{\bQ}}
\newcommand{\hbX}{\widehat{\bX}}
\newcommand{\hbZ}{\widehat{\bZ}}

\newcommand{\hbLam}{\widehat{\bLam}}

\newcommand{\hbd}{\hat{\bd}}
 
\newcommand{\hL}{\widehat{L}}

\newcommand{\barK}{\overline{K}}

\newcommand{\tilC}{\widetilde{C}}
\newcommand{\tilL}{\widetilde{L}}

\newcommand{\tilbU}{\widetilde{\bU}}

\newcommand{\tilbLam}{\widetilde{\bLam}}
\newcommand{\tbLam}{\widetilde{\bLam}}

\newcommand{\bcalW}{\mbox{\mathversion{bold}$\calW$}}
\newcommand{\hbcalW}{\widehat{\bcalW}}
\newcommand{\tbcalW}{\widetilde{\bcalW}}

\newcommand{\bcalV}{\mbox{\mathversion{bold}$\calV$}}
\newcommand{\hbcalV}{\widehat{\bcalV}}
\newcommand{\tbcalV}{\widetilde{\bcalV}}

\newcommand{\hcalG}{\widehat{\calG}}
\newcommand{\hcalH}{\widehat{\calH}}

\newcommand{\frS}{\mathfrak{S}}

\newcommand{\barbZ}{\overline{\bZ}}
\newcommand{\barbU}{\overline{\bU}}
\newcommand{\barbV}{\overline{\bV}}
\newcommand{\barbD}{\overline{\bD}}

\newcommand{\barbG}{\overline{\bG}}
\newcommand{\barbQ}{\overline{\bQ}}
\newcommand{\hbarbG}{\overline{\widehat{\bG}}}

\newcommand{\barbR}{\overline{\bR}}
\newcommand{\barcalR}{\overline{\calR}}

 \newcommand{\scrP}{\mathscr{P}}

 \newcommand{\minL}{\displaystyle \min_{l=1, ....L}\ }

\newcommand{\lowc}{\underline{c}}
\newcommand{\highc}{\bar{c}}

\newcommand{\lowcrho}{\underline{c}_{\rho}}
\newcommand{\highcrho}{\bar{c}_{\rho}}

\newcommand{\rhonl}{\rho_{n,l}}
\newcommand{\rhon}{\rho_{n}}

\long\def\ignore#1{}

\newcommand{\reals}{\mathbb{R}}

\newcommand{\upl}{^{(l)}}
\newcommand{\upm}{^{(m)}}

\newcommand{\linL}{l \in [L]}
\newcommand{\minM}{m \in [M]}

\newcommand{\kinKm}{k \in [K_m]}

\newcommand{\sinTe}{\sin\Te}
\newcommand{\tim}{\times_3}

\firstpageno{1}

\graphicspath{%
    {converted_graphics/}
    {/}
}
\begin{document}

\title{Clustering of Diverse Multiplex Networks}

\author{     \name Marianna Pensky\footnote{Corresponding Author} \email Marianna.Pensky@ucf.edu \\
      \addr Department of Mathematics\\
       University of Central Florida\\
     Orlando, FL 32816, USA     
       \AND
    \name  Yaxuan Wang \email yxwang.math@knights.ucf.edu \\
      \addr Department of Mathematics\\
       University of Central Florida\\
     Orlando, FL 32816, USA       }

\editor{}

\maketitle

\begin{abstract}%
The paper introduces the  DIverse MultiPLEx  Generalized Dot Product Graph (DIMPLE-GDPG)
network  model where all layers of the network have  the same collection of nodes and follow the 
Generalized Dot Product Graph (GDPG)  model. In addition, all layers can be partitioned into groups 
such that the layers in the same group are embedded in the same ambient subspace
but otherwise all matrices of connection probabilities can be different. 
In a common particular case, where layers of the network follow the 
Stochastic Block Model (SBM),  this setting implies that the groups of layers have common community structures but 
all matrices of block connection probabilities can be different. We refer to this version as the DIMPLE model.
While the DIMPLE-GDPG model generalizes the COmmon Subspace Independent Edge (COSIE) random graph model developed in \cite{JMLR:v22:19-558},
the DIMPLE model includes a wide variety of SBM-equipped multilayer network models as its particular cases.
In the paper, we  introduce novel algorithms for the recovery of similar groups of layers, 
for the estimation of the ambient subspaces in the groups of layers in the DIMPLE-GDPG
setting, and for the within-layer clustering in the case of the DIMPLE model. 
We study the accuracy of those algorithms, both theoretically and via computer simulations.
The advantages of the new models are demonstrated using real data examples. 
\end{abstract}

\begin{keywords}
Multiplex Network,  Stochastic Block Model, Community Detection, Spectral Clustering
\end{keywords}


 
\section{Introduction}
\label{sec:introduction}

\subsection{Multiplex network models}
\label{sec:multiplex}

Stochastic network models appear in a variety of applications, including 
genetics, proteomics, medical imaging, international relationships, brain science and many more. 
While in the early years of the field of stochastic networks, research  mainly focused  on studying a
single network, in recent years the frontier moved to investigation of collection of networks, the
so called {\it multilayer network}, which allows to study relationships between nodes
with respect to various modalities (e.g., relationships between species based on food or space),
or consists of  network data collected from different individuals (e.g., brain networks).
Although there are many different ways of modeling a multilayer network (see, e.g.,
an excellent review article of  \cite{10.1093/comnet/cnu016}), in this paper, we consider the case where
all layers have the same set of  nodes, and all the edges between nodes are drawn within  layers, i.e.,
there are no edges connecting the nodes in different layers. 
Many authors, who work  in a variety of research fields, study this particular version of a multilayer network 
(see, e.g., \cite{Aleta_2019},  \cite{JMLR:v18:16-391}, \cite{han2018multiresolution},
 \cite{Kao_2017}, \cite{macdonald2021latent} among others).
 \cite{macdonald2021latent}
called this type of multilayer network models the {\it Multiplex Network Model} and argued that it  appears 
in a variety of real life situations.

For example,    multiplex network models include brain networks 
where nodes  are associated with brain regions, and edges are drawn if 
signals in those regions exhibit some kind of similarity (\cite{Sporns_2018}).   
In this setting, the nodes are the same for each individual network, and there is no connection between 
brain regions of different individuals. Another type of multiplex networks are 
trade networks between a set of countries (see, e.g., \cite{doi:10.1038/ncomms7864}), where nodes 
and layers represent, respectively, various countries and commodities in which they are trading. 
In this case,  edges are drawn if countries trade   specific products with each other.
In this paper we consider the following model.


\subsection{DIverse MultiPLEx (DIMPLE) network  models frameworks}
\label{sec:model}

Consider an $L$-layer network on the same set of $n$ vertices $[n] = \{1,\cdots,n\}$, where
the tensor of probabilities of connections $\calP \in [0,1]^{n \times n \times L}$
is formed by layers $\bP^{(l)}$, $\linL$, that can be partitioned into   $M$  groups
with the common subspace structure or community assignment.

In this paper, we consider a multiplex network with $L$ layers 
of $M$ types,  so that there exists a label function $c: [L] \to [M]$.
We assume that the layers of the network follow the Generlized Dot Product Graph ({\bf GDPG}) model of 
\cite{GDPG}, where each group of layers is embedded in its own ambient subspace, but otherwise all matrices 
of connection probabilities can be different. Specifically, $\bP\upl$, $l \in [L]$, are given by
\be \label{eq:DIMPLE_GDPG}
\bP\upl = \bV\upm \bQ\upl (\bV\upm)^T, \quad m = c(l), \ l \in [L],\ m \in [M],
\ee
where $\bQ\upl = (\bQ\upl)^T$ and $\bV\upm$ are matrices  with orthonormal columns,  such that 
all entries of $\bP\upl$ are in $[0,1]$. 
We shall call this model the DIverse MultiPLEx  Generalized Dot Product Graph ({\bf DIMPLE-GDPG}).

In a common particular case, where layers of the network follow the 
Stochastic Block Models ({\bf SBM}),  \eqref{eq:DIMPLE_GDPG} implies that the groups of layers have common community structures but 
 matrices of block connection probabilities can be all different.
Then, the matrix of probabilities of connection in layer $l$ can be expressed as 
\be \label{eq:model}
\bP\upl = \bZ\upm \boB\upl (\bZ\upm)^T, \quad m = c(l), \ \linL,\  \minM, 
\ee 
where  $\bZ\upm$  is the clustering matrix in the layer of type $m = c(l)$
and $\boB\upl  = (\boB\upl)^T$ is a matrix of block probabilities, $\linL$.
In order to  distinguish this special case, we shall refer to \eqref{eq:model}
as simply the {\bf DIMPLE} model.

In both models, one observes the adjacency tensor $\calA \in \{0,1\}^{n \times n \times L}$ with layers $\bA\upl$ such that 
$\bA{\upl} (i,j) = \bA\upl (j,i)$ and, for $1 \leq i < j \leq n$ and $\linL$, where $\bA\upl (i,j)$ are the Bernoulli
random variables with $\PP (\bA\upl(i,j)=1)=\bP \upl (i,j)$, and they are independent from each other. 
The objective is to recover the layer clustering matrix $\bC$, as well as the community assignment 
matrices $\bZ\upm$  in the case of model \eqref{eq:model}, or the subspaces $\bV\upm$ 
in the case of model \eqref{eq:DIMPLE_GDPG}. 

Note that,  since the SBM is a particular case of the GDPG, 
\eqref{eq:model} is a particular case of \eqref{eq:DIMPLE_GDPG} (see Section~\ref{sec:between} for further explanations). 
Nevertheless, the problems associated with  \eqref{eq:DIMPLE_GDPG} and \eqref{eq:model}  are somewhat different.
While recovering matrices $\bV\upm$ is an estimation problem, finding communities in the groups of layers,
corresponding to clustering matrices $\bZ\upm$, is a clustering problem. 
For this reason, we study both models, \eqref{eq:DIMPLE_GDPG} and \eqref{eq:model}, in this paper.    
%
\\

\noindent
Our paper makes several key contributions. 

\begin{enumerate} 

\item
Our paper is the first one that considers the SBM-equipped multiplex network, where both the
probabilities of connections and the community structures can vary. In this sense,
our paper generalizes both the models,   where the community structure 
is identical in all layers, and the  ones, where there are only $M$ types of the 
matrices of the connection probabilities, so that the probability tensor has 
collections of identical layers. Those models correspond, respectively,  to $M=1$, 
and to $\boB\upl = \boB\upm$ with $m = c(l)$ in \eqref{eq:model}.

\item
Our paper generalizes the COmmon Subspace Independent Edge ({\bf COSIE}) random graph model 
of \cite{JMLR:v22:19-558}  and \cite{MinhTang_arxiv2022}, which corresponds to 
$M=1$ in \eqref{eq:DIMPLE_GDPG}.

\item 
Our paper develops a novel between-layer clustering algorithm that works for both DIMPLE and DIMPLE-GDPG 
network model and   derive  expressions for the clustering errors 
under very simple and intuitive assumptions. Our simulations confirm 
that the between-layer and the within-layer clustering 
algorithms deliver high precision in a finite parameter settings.
In addition, if $M=1$, our subspace recovery error compares favorably to 
the ones in \cite{JMLR:v22:19-558}  and \cite{MinhTang_arxiv2022}, 
due to employment of a different algorithm.

\item
Since the  DIMPLE model generalizes two types of popular   SBM-equipped
multiplex networks models, our paper  opens a gateway for testing/model selection.
In particular, one can test  whether communities persist throughout 
the layers of the network, or whether layers can be partitioned into groups for which this is true,
which is equivalent to testing the hypothesis that  $M=1$ in \eqref{eq:model}.
Alternatively, one can test the hypothesis that all matrices $\boB\upl$
in a  group of layers are the same that reduces to $\boB\upl = \boB\upm$ with $m = c(l)$ in \eqref{eq:model}.
One can test similar hypotheses in the case of  the DIMPLE-GDPG network model.

\end{enumerate}


The rest of the paper is organized as follows. 
Section~\ref{sec:rel_work} reviews related work, explains why introduction of the 
DIMPLE and the DIMPLE-GDPG models is imperative, and why  analysis of those models
requires development of new  algorithms. Following it, 
Section~\ref{sec:notations} introduces    notations,  required
for construction of the algorithms and their subsequent analysis. 
Section~\ref{sec:DIMPLE_fitting} is devoted to fitting the   DIMPLE and the DIMPLE-GDPG network models. 
In particular, Section~\ref{sec:between} proposes  a between-layer clustering algorithm
for both the   DIMPLE and the DIMPLE-GDPG models.
Section~\ref{sec:DIMPLE_GDPG} talks about estimation of invariant subspace matrices $\bV\upm$
in the groups of layers in the DIMPLE-GDPG model in \eqref{eq:DIMPLE_GDPG}.
Section~\ref{sec:within} provides within-layer clustering procedures 
in the case of  the DIMPLE network. 
Section~\ref{sec:theory} is dedicated to theoretical developments. Specifically,
Section~\ref{sec:assump} introduces assumptions that guarantee the 
between-layer clustering error rates,   the within-layer clustering error rates for the DIMPLE model and 
the subspace  fitting errors in groups of layers in the DIMPLE-GDPG model, that are derived in 
Sections~\ref{sec:error_between}, \ref{sec:error_within} and \ref{sec:error_fitting}, respectively.
Section~\ref{sec:simul_study} presents simulation studies for the DIMPLE and the DIMPLE-GDPG model.
Section~\ref{sec:real_data} provides   real data examples where algorithms developed 
in the paper are applied to  the worldwide food trading networks data and airline data.
Section~\ref{sec:discussion} concludes the paper with the discussion of its results.
Finally, Section~\ref{sec:appendix} contains proofs of the statements in the paper and also provides 
additional simulations.



\subsection{Justification of the  model and related work}
\label{sec:rel_work}

In the last few years, a number of authors studied multiplex network models.
The vast majority of the paper assumed that all layers of the network 
follow the Stochastic Block Model  (SBM). The latter is due to the fact that the 
SBM, according to \cite{Olhede14722},  provides a universal tool for description of time-independent stochastic network data.
It is also very common in applications. 
For example, \cite{Sporns_2018} argues that stochastic block models provide a powerful tool for brain studies.
In fact, in the last few years,  such models have been widely employed in brain research (see, e.g., \cite{crossley2013cognitive},
\cite{pub.1106343698}, \cite{nicolini2017community}, among others).

While the scientific community considered various types of multiplex networks in general, and the SBM-equipped 
multiplex networks in particular 
(see e.g., \cite{doi:10.1098/rsos.171747}, \cite{Kao_2017}, \cite{mercado2018power} among others),
the theoretically inclined papers in the field of statistics   mainly have been  investigating the case where
communities persist throughout all layers of the network. 
This includes studying the so called ``checker board model''  in \cite{JMLR:v21:18-155},  
where the matrices of block probabilities take only finite number of values,
and communities are the same in all layers. The tensor block models of \cite{NEURIPS2019_9be40cee} and \cite{han2021exact} 
belong to the same category. In recent years, statistics publications extended this type of research   
to the  case where community structure is preserved in all layers of the network, 
but the matrices of block connection probabilities   can take
arbitrary values (see, e.g., \cite{bhattacharyya2020general},  \cite{10.1093/biomet/asz068},  \cite{lei2021biasadjusted},
 \cite{paul2016}, \cite{paul2020}  and references therein). The authors studied  precision of community detection, 
and provided theoretical and numerical comparisons between various techniques that can be employed in this case.

In addition,  the recent years saw a substantial advancement in the latent position graphical models.
Specifically, the Random Dot Product Graph (RDPG) model of  \cite{JMLR:v18:17-448}  and the  
Generalized Dot Product Graph (GDPG) model of  \cite{GDPG}   turned out to be very flexible and useful in applications. 
In the last few years,  \cite{JMLR:v22:19-558}  
and \cite{MinhTang_arxiv2022} introduced the COmmon Subspace Independent Edge ({\bf COSIE}) random graph model 
which  extends the  RDPG and the GDPG to the multilayer setting.
 However,  COSIE postulates that the layer networks are embedded into the same 
invariant subspace, which is very similar to the assumption of persistent communities  in all layers of a multiplex network.

Nevertheless, there are many real life scenarios where the assumption, that all layers of the network
have the same communities or are embedded into the same subspace is too restrictive. 
For example, it is known that some brain disorders are associated with
changes in brain network organizations (see, e.g., \cite{Buckner2019TheBD}), and that alterations in the community structure of
the brain have been observed in several neuropsychiatric conditions, including Alzheimer disease (see, e.g., \cite{doi:10.1002/hbm.23240}),
schizophrenia (see, e.g., \cite{pub.1037745277}) and epilepsy disease (see, e.g., \cite{munsell_2015}). 
%
In this case, one would like to examine brains networks of the individuals with and without brain disorder 
to derive the differences in community structures.  
Similar situations occur when  one examines several groups of networks, often corresponding to 
subjects with different biological conditions (e.g., males/females, healthy/diseased, etc.)

One of the possible approaches here is to  assume that both, the community structures 
and the probabilities of connections in the  network layers, will be identical under the same  biological  condition 
and  dissimilar for different conditions.  
This type of setting, called the {\bf M}ixture {\bf M}ulti{\bf L}ayer {\bf S}tochastic
{\bf B}lock {\bf M}odel  ({\bf MMLSBM})  assumes that all layers 
can be partitioned into a few different types,such that  each distinct type of   
layers is equipped with its own  community structure and a unique  matrix of block connection probabilities,
and that both are identical within the same type of layers.  In the context of a GDPG-based 
multiplex network, this extension leads   directly to low-rank tensor estimation, 
the problem that received a great deal of attention in the last five years.

Specifically, if $M=1$, then the DIMPLE model \eqref{eq:model} reduces to the multiplex models in 
\cite{bhattacharyya2020general},  \cite{10.1093/biomet/asz068}, \cite{lei2021biasadjusted},
 \cite{paul2016}, \cite{paul2020} with the persistent communities, 
and it becomes the MMLSBM of \cite{Stanley2019}, \cite{TWIST-AOS2079} and \cite{fan2021alma},
if $\boB\upl$ takes only $M$ distinct values, i.e.,  $\boB\upl = \boB \upm$ for $c(l)= m$.
Similarly,  if $M=1$, the DIMPLE-GDPG model in \eqref{eq:DIMPLE_GDPG} reduces to the COSIE model in 
\cite{JMLR:v22:19-558} and  \cite{MinhTang_arxiv2022}, and it reduces to a low rank tensor estimation
of \cite{JMLR:AZhang21} and \cite{ZhangXiaIEEE2018} if all matrices $\bQ\upl$ are identical within a group of layers.

In essence, the conclusion of the discussion above is that so far authors considered two complementary types of settings
for multiplex networks. In the first of them, all layers of the network are embedded into the same subspaces
 in the case of the   GDPG, or have the same communities if the layers of the network are equipped with SBMs. 
In the second one, the layers may be embedded into different subspaces, but the tensor of connection probabilities has a low rank,
which reduces to MMLSBM if   layers follow the SBM.

Hence, the natural generalization of those two scenarios would be the setting, where the layers of 
the network can be partitioned into groups, each with the distinct subspace or community structure.
Such multiplex network can be viewed as a concatenation of several multiplex networks that follow
COSIE model or Stochastic Block Models with persistent community structure. On the other hand, 
such networks will reduce to a low rank tensor or the MMLSBM if networks in the group of layers have 
identical probabilities of connections.

We feel that the above extension is imperative for a variety of reasons. 
As one can easily see, the existing models are complementary in nature and are usually adopted without 
any consideration of the alternatives. The DIMPLE-GDPG and the DIMPLE models allow  to forgo this choice.
They also open the gate for testing this alternatives and adopting the   one which better fits the data. 
Our real data examples show that  in real life situations the DIMPLE or the DIMPLE-GDPG model provides a 
better summary of data than the MMLSBM.

The new DIMPLE-GDPG model requires development of new algorithms, since the  
probability tensor $\calP$ associated with the DIMPLE-GDPG model in \eqref{eq:DIMPLE_GDPG} does not 
have a low  rank, due to the fact that all matrices $\bQ\upl$ are different. For this reason, techniques and 
theoretical assessments developed for low rank tensors do not work in the case of the DIMPLE-GDPG model. 
Similarly, since the  matrices of the block connection probabilities 
take different values in each of the layers, techniques  employed in  \cite{TWIST-AOS2079} and \cite{fan2021alma} 
cannot be applied in the new environment of DIMPLE. 

Indeed, the TWIST algorithm of \cite{TWIST-AOS2079}  
is based on the alternating regularized low rank approximations of  the adjacency tensor, which 
 relies on the fact that the tensor of connection probabilities is truly low rank in the case of MMLSBM.
This, however, is not  true for the DIMPLE model, where the matrices of block connection probabilities 
vary from layer to layer. On the other hand, the ALMA algorithm of \cite{fan2021alma} exploits the fact 
that the  matrices of connection probabilities   are identical in the groups of layers with the same community structures. 
This is no longer the case in the environment of the DIMPLE model, where matrices of connection probabilities  are
all different for different layers. Specifically, Section~\ref{sec:DINMPLE-MMLSBM} compares the MMLSBM and the DIMPLE model 
introduced in this paper and shows that, while algorithms designed for the DIMPLE model
work well for the MMLSBM, the algorithms designed for the MMLSBM display poor performance 
if data are generated according to the DIMPLE model.



\subsection{Notations}
\label{sec:notations}

For any integer $n$, we denote $[n] = \{1, ..., n\}$. 
We   denote tensors by calligraphy letters and matrices by bold letters. 
Denote by $\calM_{N,K}$   the set of the {\it clustering }  matrices for $N$ objects partitioned into $K$ groups
\bes
\calM_{N,K} = \lfi \bX \in \{0,1\}^{N \times K},\quad \bX \bone = \bone, \quad \bX^T \bone  \neq \bzero \rfi,
\ees
where $\bX\in \calM_{N,K}$ are such that $\bX_{i,j} = 1$ if node $i$ is in cluster $j$ and 
and $\bX_{i,j}  = 0$  otherwise.
For any matrix $\bX$, denote the Frobenius, the infinity and the operator norm   by
$\|\bX\|_F$, $\|\bX\|_{\infty}$  and $\|\bbX\|$, respectively, and its $r$-th largest
singular value by $\sig_r(\bX)$. Let $\displaystyle \|\bX\|_{2 \to \infty} = \sup_{\|z\|=1} \|\bX z\|_{\infty}$.

The column $j$ and the row $i$ of a matrix $\bQ$ are denoted by  $\bQ (:,j)$ and $\bQ(i,:)$, respectively.
Denote the identity and the zero matrix of size $K$ by, respectively, $\bI_K$ and $\bzero_K$ 
(where $K$ is omitted when this does not cause ambiguity).  
Denote
\be \label{eq:orth}
\calO_{n,K} = \lfi \bX \ \in \RR^{n \times K}:\ \  \bX^T \bX = \bI_K \rfi, \quad \calO_n=\calO_{n,n}.
\ee
Let   $\vect(\bX)$ be the vector obtained from matrix $\bX$ by sequentially stacking its columns. Denote  
by $\bX \otimes \bY$ the Kronecker product of matrices $\bX$ and $\bY$. Denote $n$-dimensional vector with unit components 
by $\bone_n$. Denote diagonal of a matrix $\bA$ by $\diag(\bA)$. Also, denote the $M$-dimensional diagonal matrix with 
$a_1, ..., a_M$ on the diagonal by $\diag(a_1, ..., a_M)$.

For any  matrix $\bX \in \RR^{n_1 \times n_2}$, denote its projection on the nearest rank $K$ matrix by $\Pi_K (\bX)$,
that is, if  $\sig_k$ are the singular values,  and $u_k$ and $v_k$ are the left and the right singular vectors of $\bX$, 
$k=1, ...,r$,  then
\bes
\bX = \sum_{k=1}^r \sig_k u_k v_k^T \quad  \Longrightarrow \quad  \Pi_K (\bX) = \sum_{k=1}^{\min(r,K)}\, \sig_k u_k v_k^T. 
\ees
For any matrices $\bX \in \RR^{n_1 \times n_2}$ and  $\bU \in \bO_{n_1,K}$, $K \leq  n_1$,   projection of $\bX$  on the column space of $\bU$ 
and on its orthogonal space are defined, respectively, as 
\bes
\Pi_{\bU} (\bX) = \bU \bU^T \bX, \quad \Pi_{\bU_{\bot}} (\bX) = (\bI - \Pi_{\bU})  \bX. 
\ees
Following \cite{Kolda09tensordecompositions}, we define the following tensor operations.
For any tensor $\calX\in \reals^{n_1\times n_2\times n_3}$ and a matrix $\bA\in\reals^{m \times  n_3}$,
their product $\calX\times_3\bA$ along dimension 3 is a tensor in $\reals^{n_1 \times n_2\times m}$ with elements
\bes
[\calX\times_3\bA](i_1,i_2,j) = \sum_{i_3=1}^{n_3} \bA(j,i_3)\calX(i_1,i_2,i_3), \quad j=1, ..., m.
\ees
If $\calY\in\reals^{m\times n_2\times n_3}$ is another tensor,  the product between tensors $\calX$ and $\calY$
along dimensions (2,3), denoted by $\calX\times_{2,3}\calY$, is a matrix in $\reals^{n_1\times m}$ with elements
\bes
[\calX\times_{2,3}\calY](i_1,i_2)=\sum_{j_2=1}^{n_2}\sum_{j_3=1}^{n_3}\calX(i_1,j_2,j_3)\calY(i_2,j_2,j_3),
\quad i_1 = 1, ..., n_1, \ i_2 = 1, ..., m.
\ees
The mode-3 matricization of tensor $\calX \in \reals^{n_1\times n_2\times n_3}$ is a matrix $\mathscr{M}_3(\calX) = \bX \in \RR^{n_3 \times (n_1 n_2)}$ 
with rows $\bX (i,:) = [\vect(\calX(:,:,i))]^T$. Please, see \cite{Kolda09tensordecompositions} for  a more extensive 
discussion of tensor operations and their properties.

We use the $\sinTe$ distances to measure the separation between two subspaces with orthonormal 
bases  $\bU \in \calO_{n,K}$ and $\widetilde{\bU} \in \calO_{n,K}$, respectively. 
Suppose the singular values of $\bU^T \widetilde{\bU}$ are 
$\sig_1 \geq \sig_2 \geq ... \geq \sig_K>0$. Then  
\bes 
\Te(\bU,\widetilde{\bU}) = \diag \lkr \cos^{-1}(\sig_1), ..., \cos^{-1}(\sig_K) \rkr
\ees
are the principle angles. Quantitative measures of the distance between the
column spaces of $\bU$ and $\widetilde{\bU}$ are then 
\be  \label{eq:sinTheta}
 \left\| \sin \Te(\bU,\widetilde{\bU}) \right\| = \sqrt{1 - \sig_{\min}^2 (\bU^T \widetilde{\bU})}  
\quad \mbox{and} \quad
 \left\| \sin \Te(\bU,\widetilde{\bU}) \right\|_F = \sqrt{K - \|\bU^T \widetilde{\bU})\|^2_F} 
\ee  
Some convenient characterizations of those distances can be found in 
Section 8.1 of \cite{10.1214/17-AOS1541}. 

Finally, we shall use $C$ for a generic positive constant  that can take different values and
is independent of $L$, $n$, $M$, $K$ and graph densities.


 
\section{Fitting the DIMPLE and the DIMPLE-GDPG models }
\label{sec:DIMPLE_fitting}

In this paper, we consider a multiplex network with $L$ layers 
of $M$ types, where $L_m$ is the number of layers of type $m$, $\minM$. Let $\bC \in \calM(L,M)$ be  the layer clustering matrix.
A layer of type $m$  has an ambient dimension  $K_m$. In the case of model \eqref{eq:model},
a layer of type $m$  has $K_m$ communities, 
and $n_{k,m}$ is the number of nodes of type $k$ in the layer of type $m$, $\kinKm$,   $\minM$,
so that
\be \label{eq:diagonal}
\bD_z\upm = (\bZ\upm)^T  \bZ\upm  = \diag(n_{1,m}, ..., n_{K_m,m}).
\ee


\subsection{Between-layer clustering}
\label{sec:between}  

First, we show that model  \eqref{eq:model} is a particular case of model \eqref{eq:DIMPLE_GDPG}.
Indeed, denote $\bU_z\upm = \bZ\upm \lkr \bD_z\upm \rkr^{-1/2}$, where  matrices $\bD_z\upm$ are defined in \eqref{eq:diagonal}.
Since $\bU_z\upm   \in \calO_{n,K_m}$,  matrices $\bP\upl$ in \eqref{eq:model} can be written as 
\be  \label{eq:expans_SBM}   
\bP\upl =  \bU_z\upm \boB_D\upl (\bU_z\upm)^T, \quad 
\boB_D\upl =   \sqrt{\bD_z\upm}\,   \boB\upl \, \sqrt{\bD_z\upm}  
\ee 
Therefore, \eqref{eq:model} is a particular case of \eqref{eq:DIMPLE_GDPG} with $\bV\upm = \bU_z\upm$
and $\bQ\upl =   \boB_D\upl$. For this reason, we are going to 
cluster groups of layers in the more general setting \eqref{eq:DIMPLE_GDPG}  of DIMPLE-GDPG.

In order to find  the clustering matrix $\bC$,  observe that matrices $\bP\upl$ in \eqref{eq:DIMPLE_GDPG} 
can be written as 
\be  \label{eq:expans1}
\bP\upl =  \bV\upm   \bO_Q\upl \bS_Q\upl (\bO_Q\upl)^T (\bV\upm)^T, \quad \linL
\ee 
where 
\be    \label{eq:Ql}   
\bQ\upl =  \bO_Q\upl \bS_Q\upl (\bO_Q\upl)^T, \quad \linL,
\ee  
is the  singular value decomposition ({\bf SVD}) of $\bQ\upl$ with $\bO_Q\upl \in \calO_{n,K_m}$, $m = c(l)$, and diagonal matrix 
$\bS_Q\upl$.
In order to extract common information from matrices  $\bP\upl$, we consider the SVD of  $\bP\upl$
\be \label{eq:svd1} 
\bP\upl = \bU_{P,l} \bLam_{P,l} (\bU_{P,l})^T, \quad \bU_{P,l} \in \calO_{n,K_m},\ \linL,\ m = c(l)
\ee 
and relate it to expansion \eqref{eq:expans1}. 
If, as we assume later, matrices $\bQ\upl$ are of full rank, then 
$\bO_Q\upl \in \calO_{K_m}$, so that $\bO_Q\upl (\bO_Q\upl)^T = (\bO_Q\upl)^T \bO_Q\upl = \bI_{K_m}$,
$m = c(l)$. Therefore,   $\bV\upm   \bO_Q\upl  \in  \calO_{n,K_m}$, and  expansion \eqref{eq:expans1} is 
just another way of writing the SVD of $\bP\upl$. Hence, up to the $K_m$-dimensional rotation $\bO_Q\upl$,
matrices $\bV\upm$ and $\bU_{P,l}$ are equal to each other when $c(l)=m$ .

%
\begin{algorithm} [t] 
\caption{\ The between-layer clustering}
\label{alg:between}
\begin{flushleft} 
{\bf Input:} Adjacency tensor $\calA \in \{0,1\}^{n \times n \times L}$; number of groups of layers $M$;  
ambient dimension $K^{(l)}$ of each layer $\linL$; parameter $\eps$ \\
{\bf Output:} Estimated clustering matrix $\hbC \in \calM_{L,M}$ \\
{\bf Steps:}\\
{\bf 1:} Find the  SVDs\   $\Pi_{K^{(l)}} (\bA\upl) = \hbU_{A,l} \hbLam_{P,l} (\hbU_{A,l})^T$, $\linL$ \\
{\bf 2:} Form matrix $\hbTe \in \RR^{n^2 \times L}$ with columns $\hbTe(:,l) = \vect(\hbU_{A,l} (\hbU_{A,l})^T)$\\
{\bf 3:} Construct the SVD of $\hbTe$ using \eqref{eq:svd_hbTe} and obtain matrix $\hbcalW = \tbcalW(:, 1:M) \in \calO_{L,M}$\\
{\bf 4:}  Cluster  $L$ rows of  $\hbcalW$ into $M$ clusters using   $(1+\eps)$-approximate $K$-means clustering. Obtain 
estimated clustering matrix $\hbC$   
\end{flushleft} 
\end{algorithm}
%


Since matrices $\bO_Q\upl$ are unknown, we introduce   alternatives to $\bU_{P,l}$:
\be \label{eq:main_rel}
\bU_{P,l} (\bU_{P,l})^T = \bV\upm   \bO_Q\upl (\bV\upm   \bO_Q\upl)^T   = \bV\upm (\bV\upm)^T, \quad m = c(l),
\ee
which depend  on $l$ only via $m = c(l)$ and are uniquely defined for $\linL$.
The latter implies that the between-layer clustering can be based on  the
matrices $\bU_{P,l} (\bU_{P,l})^T$, $\linL$, or rather on their vectorized versions. 
Denote
\be \label{eq:bC_structure}   
\bD_{c} = \bC^T \bC = \diag(L_1, ...,L_M), \quad
\bW = \bC (\bD_{c})^{-1/2}  \in \calO_{L,M}
\ee 
Consider matrices $\bPsi \in \RR^{n^2 \times M}$ and $\bTe \in \RR^{n^2 \times L}$ with respective  columns 
\bes
\bPsi (:,m) = \vect(\bV\upm (\bV\upm)^T), \quad 
 \bTe (:,l) = \vect \lkr \bV^{(c(l))} (\bV^{(c(l))})^T \rkr = \vect(\bU_{P,l} (\bU_{P,l})^T),\ 
\ees  
where $\minM$,  $\linL$.
It is easy to see that 
\be \label{eq:rel1}
\bTe = \bPsi \bC^T, \quad \bPsi = \bTe \bC \bD_c^{-1},
\ee 
 so that clustering assignment can be recovered by spectral clustering of  columns of 
an estimated version of matrix $\bTe$.

For this purpose, consider layers $\bA\upl = \calA(:,:,l)$ of the adjacency tensor $\calA$
and construct the  SVDs of their rank $K_m$ projections $\Pi_{K_m} (\bA\upl)$: 
\be  \label{eq:svd_A_l}
\Pi_{K_m} (\bA\upl) = \hbU_{A,l} \hbLam_{P,l} (\hbU_{A,l})^T, \quad \hbU_{A,l} \in \calO_{n,K_m},\quad m = c(l),\ \linL.
\ee 
Then,  replace matrix $\bTe$ by its proxy  $\hbTe$ with columns
$\hbTe (:,l) = \vect(\hbU_{A,l} (\hbU_{A,l})^T)$. The major difference between $\bTe$ and $\hbTe$,
however, is that, under assumptions in Section~\ref{sec:assump}, $\rank(\bTe) = M$ while, in general, $\rank(\hbTe) = L>> M$.
If the SVD of $\hbTe$ is
\be \label{eq:svd_hbTe}
\hbTe = \tbcalV \tbLam  \tbcalW, \quad \tbcalV \in \calO_{n^2,L},\quad  \tbcalW \in \calO_L,
\ee
then, we can form reduced matrices
\be \label{eq:reduced_SVD_hbTe}
\hbcalV = \tbcalV(:, 1:M) \in \calO_{n^2,M}, \quad \hbcalW = \tbcalW(:, 1:M) \in \calO_{L,M}, 
\ee
%
%
and  apply clustering to the rows of  $\hbcalW$  rather than to the rows of $\tbcalW$.
The latter results in Algorithm~\ref{alg:between}. We use   $(1+\eps)$-approximate $K$-means clustering
to obtain the final clustering assignments. There exist efficient algorithms 
for solving the $(1+\epsilon)-$approximate $K$-means problem  
(see, e.g., \cite{1366265}). We denote 
\be \label{eq:hatW}
\hbD_c = \hbC^T \hbC,\quad \hbW = \hbC \hbD_c^{-1/2} \in \calO_{L,M}
\ee
Observe that clustering procedure above relies on the knowledge of the ambient dimension $K_m$, 
which is associated with the unknown group membership $m = c(l)$. Instead of assuming 
that $K_m$ are known, as it is done in \cite{TWIST-AOS2079} and \cite{fan2021alma},
we assume that one knows the ambient dimension $K^{(l)}$ of the GDPG in every layer $\linL$ 
of the network. This is a very common assumption and is imposed in almost every paper 
that studies latent position or block model equipped networks 
(see, e.g.,  \cite{JMLR:v18:17-448}, \cite{GDPG}, \cite{gao2018},  \cite{Gao:2017:AOM:3122009.3153016}).
In this case, one can replace $K_m$ in \eqref{eq:svd_A_l}  by $K^{(l)}$. 
We further discuss this issue in Remark~\ref{rem:unknown_Km}.

\begin{rem} \label{rem:unknownM}
{\bf Unknown number of layers. }
{\rm
While Algorithm~\ref{alg:between} assumes $M$ to be known, in many practical situations this is not true, and the 
value of $M$  has to be discovered from data.
Identifying the number of clusters is a common  issue in data clustering, and it is 
a separate problem from the process of actually solving the clustering problem
with a known number of clusters.  
A common method for finding the number of clusters is the so called ``elbow'' method 
that looks at the fraction of the variance 
explained as a function of the number of clusters. The method is based on the
idea that one should choose the smallest number of clusters, such that adding
another cluster does not significantly improve fitting of the data by a model.
There are many ways to determine the ``elbow''. For example, one can base its detection on 
evaluation of the clustering error in terms of an objective function, as in, e.g., \cite{Zhang2012}.
Another possibility is to  monitor  the eigenvalues of the non-backtracking matrix or the 
Bethe Hessian matrix, as it is done in~\cite{Le2015EstimatingTN}. 
One can also  employ a simple technique of checking the eigen-gaps of the matrix 
$\tilbLam$ in \eqref{eq:svd_hbTe}, as it has been discussed in \cite{vonLuxburg2007}, 
or use a scree plot as it is done in \cite{ZHU2006918}. 
}
\end{rem}

\begin{rem} \label{rem:unknown_Km}
{\bf  Unknown  ambient dimensions.\ }
{\rm
In this paper, for the purpose of methodological developments, we assume that the 
ambient dimension $K^{(l)}$ of each layer of the network is known (which corresponds 
to the known number of communities in the case of the DIMPLE model). This is a common assumption,
and everything in the Remark~\ref{rem:unknownM} can also be applied to this case. 
Here, $K^{(l)}= K_m$  with $m = c(l)$. One can, of course, can assume that the values of $K_m$,
$\minM$, are known. However, since group labels are interchangeable, in the case of non-identical subspace dimensions 
(numbers of communities), it is hard to choose, which of the values corresponds to which of the  groups.  
This is actually the reason why \cite{TWIST-AOS2079} and \cite{fan2021alma}, who imposed this assumption,
used it only in theory, while their simulations and real data examples are all restricted to the case 
of equal number of communities in all layers $K_m = K$, $\minM$.  On the contrary, knowledge of $K\upl$ allows one to deal 
with different ambient dimensions (number of communities) in the groups of layers in simulations and real data examples.

Of course, if $K_m$ are all different, e.g., $M=3$, $K_1 = 2$, $K_2=3$ and $K_3=4$,
this seems to imply that one can use this information for clustering of layers. 
However, this is not true in general. Also, in practice, the values of $K^{(l)}$ are estimated,
so precision of the clustering procedure based entirely on the ambient dimensions of layers 
is questionable at best.
} 
\end{rem}


\subsection{Fitting invariant subspaces in groups of layers in the DIMPLE-GDPG model}
\label{sec:DIMPLE_GDPG}

If we knew the true clustering matrix $\bC$ and the true probability tensor  
$\calP \in \RR^{n \times n \times L}$ with layers $\bP\upl$ given by \eqref{eq:DIMPLE_GDPG},
then  we could average  layers with identical subspace structures.  
Precision of estimating $\bV\upm$, however, depends on whether the eigenvalues of $\bQ\upl$ with $c(l) =m$  add up.  
Since the latter is not guaranteed, one can alternatively add the squares   $\bG\upl= (\bP\upl)^2$, obtaining 
\bes 
\sum_{c(l) = m} \bG\upl =   \sum_{c(l) = m} (\bP\upl)^2 =  \sum_{c(l) = m} \bV\upm \, (\bQ\upl)^2  \, (\bV\upm)^T, \quad \minM
\ees
In this case, the eigenvalues of $(\bQ\upl)^2$ are all positive which ensures successful recovery of matrices $\bV\upm$.

%
%
\begin{algorithm} [t] 
\caption{\ Estimating invariant subspaces}
\label{alg:subspace_est}
\begin{flushleft} 
{\bf Input:} Adjacency tensor $\calA \in \{0,1\}^{n \times n \times L}$; number of groups of layers $M$;
ambient  dimensions $K_m$, $\minM$, of each group of layers;
estimated   clustering matrix $\hbC \in \calM_{L,M}$  \\  
{\bf Output:} Estimated invariant subspaces $\hbV\upm$, $\minM$ \\
{\bf Steps:}\\
{\bf 1:} Construct tensor $\hcalG$ with layers $\hbG\upl$ given by \eqref{eq:calhbG}, $\linL$\\
{\bf 2:} Construct tensor $\hcalH$ using formula \eqref{eq:est_tensors_alt} \\
{\bf 3:} Construct the SVDs of layers $\hbH\upm = \tilbU_{\hbH}\upm \hbLam_{\hbH}\upm (\tilbU_{\hbH}\upm)^T$,  
 $\minM$\\
{\bf 4:} Find $\hbV\upm = \tilbU_{\hbH}\upm (:,1:K_m) = \Pi_{K_m} (\tilbU_{\hbH}\upm)$, 
 $\minM$\\
\end{flushleft} 
\end{algorithm}
%
%

Note that, however, $(\bA\upl)^2$ is not an unbiased estimator of $(\bP\upl)^2$.
Indeed, while $\EE ((\bA\upl)^2)_{i,j} = ((\bP\upl)^2)_{i,j}$  for $i \neq j$, for the diagonal elements, one has 
\bes
\EE ((\bA\upl)^2)_{i,i} = (\bP\upl)^2_{i,i} + \sum_j \lkv (\bP\upl)_{i,j} - (\bP\upl)^2_{i,j}\rkv.
\ees
Therefore, following  \cite{lei2021biasadjusted}, we evaluate the  degree vector  
$\hbd\upl = \bA\upl \bone_n$  and form diagonal matrices $\diag(\hbd\upl)$ with vectors $\hbd\upl$
on the diagonals.
We construct a tensor $\hcalG \in \RR^{n \times n \times L}$ with layers 
$\hbG\upl = \widehat{\calG}(:,:,l)$ of the form
\be \label{eq:calhbG}
\hbG\upl =   \lkr \bA\upl \rkr^2 - \diag(\hbd\upl), \quad \linL
\ee
Subsequently, we combine layers of the same types, obtaining tensor $\hcalH \in \RR^{n \times n \times M}$   
\be \label{eq:est_tensors_alt}
\hcalH = \hcalG \times_3 \hbW^T, 
\ee 
where $\hbW$ is defined in \eqref{eq:hatW}. After that, $\bV\upm$, $\minM$,
can be estimated using the SVD. The procedure is described in Algorithm~\ref{alg:subspace_est}.

\begin{rem} \label{rem:average_est}
{\bf  Estimating invariant subspaces by averaging adjacency matrices.\ }
{\rm
If one knew that all matrices $\bQ\upl$, $\linL$, in \eqref{eq:DIMPLE_GDPG} have
only positive  eigenvalues, then estimation of invariant subspaces $\bV\upm$
could have been done by averaging adjacency matrices of the graphs, since
\bes 
\sum_{c(l) = m} \bP\upl =  \bV\upm \,  \lkr \sum_{c(l) = m}  \bQ\upl  \rkr \,  (\bV\upm)^T, \quad \minM
\ees
Indeed, the accuracy of spectral clustering relies on the relationship between the ratio
of the largest and the smallest nonzero eigenvalues.  The largest eigenvalues
of matrices $\bP\upl$ are  always positive due to the Perron-Frobenius theorem (see, e.g.,  \cite{Rao_Rao_1998}) 
and, hence, add up. However,  the same may not be true for the smallest  nonzero eigenvalues 
that can be positive or negative, so that their sum may not be large enough.
In this situation,  in the case of one-group ($M=1$) SBM-equipped multilayer network, 
simulation studies in \cite{paul2020} show that averaging of the adjacency matrices may not 
lead to improved precision of community detection in groups of layers. 
Furthermore, in the earlier version of our paper (\cite{https://doi.org/10.48550/arxiv.2110.05308}, ArXiv Version 2), 
 we studied averaging of the adjacency matrices in the DIMPLE model  under the 
assumption that all eigenvalues of matrices $\bP\upl$ are nonnegative. 
However, even in the presence of this assumption, averaging of adjacency matrices 
does not substantially improve the accuracy in comparison with the bias-adjusted spectral clustering 
in Algorithm~\ref{alg:subspace_est},  
while performing significantly worse when this assumption does not hold.
For this reason, we shall avoid presentation of this algorithm in our exposition.
} 
\end{rem}


\subsection{Within-layer clustering in the DIMPLE multiplex network}
\label{sec:within}

After the matrices $\bV\upm$ have been estimated, one can find clustering matrices $\bZ\upm$ 
in \eqref{eq:model} by approximate $K$-means  clustering. 
Indeed, up to a rotation, $\bV\upm$ is equal to $\bU_z\upm = \bZ\upm (\bD_z\upm)^{-1/2}$,
where $\bZ\upm$ is the clustering matrix of the layer $m$. Hence, there are only $K_m$
distinct rows in the matrix $\bV\upm$, and clustering assignment can be obtain using 
Algorithm~\ref{alg:within}.

%
\begin{algorithm} [t] 
\caption{\ The within-layer clustering}
\label{alg:within}
\begin{flushleft} 
{\bf Input:} Adjacency tensor $\calA \in \{0,1\}^{n \times n \times L}$; number of groups of layers $M$;  number of communities $K_m$, $\minM$;
estimated   clustering matrix $\hbC \in \calM_{L,M}$; parameter $\eps$ \\  
{\bf Output:} Estimated community assignments $\hbZ\upm \in \calM_{n,K_m}$, $\minM$ \\
{\bf Steps:}\\
{\bf 1:} Construct tensor $\hcalG$ with layers $\hbG\upl$ given by \eqref{eq:calhbG}, $\linL$\\
{\bf 2:} Construct tensor $\hcalH$ using formula \eqref{eq:est_tensors_alt} \\
{\bf 3:} Construct the SVDs of layers $\hbH\upm = \tilbU_{\hbH}\upm \hbLam_{\hbH}\upm (\tilbU_{\hbH}\upm)^T$,  
 $\minM$\\
{\bf 4:} Find $\hbV\upm = \tilbU_{\hbH}\upm (:,1:K_m) = \Pi_{K_m} (\tilbU_{\hbH}\upm)$, 
 $\minM$\\
{\bf 5:}  Cluster    rows of  $\hbV\upm$ into $K_m$ clusters using  $(1+\eps)$-approximate $K$-means clustering. Obtain 
clustering matrices $\hbZ\upm$, $\minM$   \\
\end{flushleft} 
\end{algorithm}
%




\section{Theoretical analysis}
\label{sec:theory}

In this section, we study the  between-layer clustering error rates of the Algorithm~\ref{alg:between},
 the error of estimation of invariant subspaces for the DIMPLE-GDPG model  of Algorithm~\ref{alg:subspace_est}, 
and  the within-layer clustering error rates  of Algorithm~\ref{alg:within}.
Since the clustering is unique only up to a permutation of cluster labels, 
denote the set of $K$-dimensional permutation functions of $[K]$ by $\aleph(K)$ and the set of $K \times K$ permutation matrices by 
$\mathfrak{F} (K)$. 
The misclassification error rate of the between-layer clustering is then given by
\be \label{eq:err_betw_def}
R_{BL} = (2\,L)^{-1}\ \min_{\scrP \in \mathfrak{F}  (M)}\  \|\hbC  - \bC \, \scrP\|^2_F.
\ee
Similarly, the  local community detection error  in the layer of type $m$ is 
\be \label{eq:err_wihin_def}
R_{WL} (m) = (2n)^{-1} \ \min_{\scrP_m \in \mathfrak{F}  (K_m)}\  \|\hbZ\upm  - \bZ\upm \, \scrP_m \|^2_F,\ \quad \minM.
\ee
Note that, since the numbering of layers is defined also up to a permutation,
the errors $R_{WL} (1)$, ..., $R_{WL} (M)$ should be minimized over the set of permutations $\aleph(M)$.
The average  error rate of the within-layer clustering is then given by
\be \label{eq:err_within_ave}
R_{WL}  =   \frac{1}{M} \ \min_{\aleph(M)}\ \sum_{m=1}^M R_{WL} (m) = \frac{1}{2\, M\,n} \ \min_{\aleph(M)}\ 
\sum_{m=1}^M  \lkv \min_{\scrP_m \in \mathfrak{F}  (K_m)}\  \|\hbZ\upm  - \bZ\upm \, \scrP_m \|^2_F \rkv
\ee
We shall measure the differences between the true and the estimated  subspace bases matrices $\bV\upm$ and $\hbV\upm$ 
using the average $\sin\Te$ distances defined in \eqref{eq:sinTheta}. Here, again  we need to 
seek the minimum over permutations of labels. We measure the errors as $R_{S,max}$ and $R_{S,ave}$ where
\be \label{eq:err_subspace_max}
R_{S,max}    =   \min_{\aleph(M)}\ \max_{\minM} \, 
 \left\| \sin \Te \lkr \bV\upm, \hbV^{(\aleph(m))} \rkr \right\|_F
\ee 
\be \label{eq:err_subspace_ave}
R_{S,ave}    = \frac{1}{M} \ \min_{\aleph(M)}\ \sum_{m=1}^M \, 
 \left\| \sin \Te \lkr \bV\upm, \hbV^{(\aleph(m))} \rkr \right\|_F^2 
\ee


\subsection{Assumptions}
\label{sec:assump}

In order the layers are identifiable,  we assume that matrices 
$\bV\upm$ in \eqref{eq:DIMPLE_GDPG} or  $\bZ\upm$ in \eqref{eq:model}  
correspond to different linear subspaces for different values of $m$.
Furthermore, the performances  of Algorithms~\ref{alg:subspace_est} and  \ref{alg:within}
depend on the success of the between-layer clustering in Algorithm~\ref{alg:between}, which, in turn,
relies on the fact that matrices   $\bV\upm (\bV\upm)^T$ in \eqref{eq:DIMPLE_GDPG} or $\bZ\upm (\bZ\upm)^T$
in \eqref{eq:model}, $\minM$, are not too similar to each other for different values of $m$. 
%

For the between layer clustering errors and the accuracy of the subspaces recovery,
we   develop our theory for the general case of the DIMPLE-GDPG model \eqref{eq:DIMPLE_GDPG}. 
Subsequently, we  derive the within-layer clustering errors for the DIMPLE model \eqref{eq:model}.
Denote
\be \label{eq:barKm}
\barK  = \frac{1}{M}\, \sum_{m=1}^M K_m, \quad K = \max_{\minM} \, K_m
\ee
Consider matrix $\barbZ \in \RR^{n \times M \barK}$, which is obtained as horizontal concatenation 
of matrices $\bV\upm \in \RR^{n \times K_m}$, $\minM$. 
Let the SVD of $\barbZ$ be
\be \label{eq:total_basis}
\barbZ = [\bV^{(1)}|...| \bV^{(M)}] = \barbU\ \barbD\ \barbV^T, \quad \barbU \in \calO_{n,r}, 
\barbV \in \calO_{M \barK,r}, \quad r \geq M+1
\ee
Here, $r$ is the rank of $\barbZ$, and $\barbD$ is an $r$-dimensional diagonal matrix. 
In the case of the DIMPLE model \eqref{eq:model}, one has  $\barbZ = [\bU_z^{(1)}|...| \bU_z^{(M)}]$.
Since matrices  $\bV\upm$   represent different subspaces, one has $M+1 \leq r < n$. 
\\

We impose the following assumptions.
\\

\noindent
{\bf A1. } Clusters of layers   are balanced, so that there exist absolute positive constants 
$C_K$, $\lowc$ and $\highc$ such that
\be \label{eq:LmKm}
C_K K \leq K_m \leq K, \quad \lowc L/M \leq L_m \leq \highc  L/M, \quad \minM
\ee 
where $L_m$ is the number of networks in the layer of type $m$. In the case of the DIMPLE model~\eqref{eq:model},
local communities  are balanced, so that 
\bes
\lowc n/K \leq  n_{k,m} \leq  \highc n/K, \quad \kinKm,  \minM
\ees
where  $n_{k,m}$  is the number of nodes in the $k$-th community in the layer of type $m$.
\\

\noindent
{\bf A2. } For some absolute constant $\kappa_0$, one has
$\sig_1(\barbD) \leq \kappa_0 \sig_r (\barbD)$ in  \eqref{eq:total_basis}.
\\

\noindent
{\bf A3. }  The layers  $\bP\upl$ of the probability tensor $\calP$ in \eqref{eq:DIMPLE_GDPG} are such that, 
for   some  absolute constant $C_\rho$
\be \label{eq:bQl}  
\bP\upl = \rhonl\, \bP_0\upl, \ \ \|\bP_0\upl\|_{\infty} =1, \quad 
\min_{\linL} \rhonl \geq C_\rho\, n^{-1}\, \log n,  
\quad \linL 
\ee
In the case of the DIMPLE model \eqref{eq:model}, \eqref{eq:bQl} reduces to 
$\boB\upl = \rhonl\, \boB_0\upl,$ $\|\boB_0\upl\|_{\infty} =1$.
\\

\noindent
{\bf A4. }  Matrices $\bQ\upl$ in \eqref{eq:DIMPLE_GDPG}  are such that, 
for some absolute constant $C_\lam \in (0,1)$, one has
\be \label{eq:minbQl}  
\minL \lkv \sig_{K_m}\lkr \bQ\upl \rkr/\sig_1 \lkr \bQ\upl\rkr \rkv \geq C_\lam,\quad m=c(l). 
\ee
In the case of the DIMPLE model, \eqref{eq:minbQl} appears as 
$\displaystyle  \min_{\linL} [\sig_{K_m}(\boB_0\upl)/\sig_1(\boB_0\upl)] \geq C_\lam$ for $m=c(l)$.
\\

\noindent
{\bf A5. } There exist absolute constants $\lowc_\rho$ and $\highc_\rho$ such that 
\be \label{eq:rho_n}
\lowc_{\rho}\, \rho_n \leq \rhonl \leq \highc_{\rho}\, \rho_n \quad \mbox{with} \quad
\rhon = L^{-1} \, \sum_{l=1}^L \rhonl
\ee
\\

\noindent
{\bf A6. }  For some absolute constant $C_{0,P}$ one has 
\be \label{eq:AssA6new}
\| \bP_0\upl\|^2_F \geq  C_{0,P}^2\, K^{-1}\,    n^2  
\ee
\\

Assumptions above are  very common and are present in many other network papers.
Specifically, Assumption~{\bf A1} is identical to  Assumptions~{\bf A3} and {\bf A4}  in \cite{TWIST-AOS2079}, 
or Assumption~{\bf A3}   in \cite{fan2021alma}.
Assumption~{\bf A2} is identical to Assumption~{\bf A2}  in \cite{TWIST-AOS2079}.
Assumption~{\bf A3}  is present in majority of papers that study community detection in 
individual networks (see, e.g. \cite{lei2015}). It   is required here since we rely on similarity of the sets of 
eigenvectors in the groups of   similar layers, and, hence, need the sample eigenvectors to converge to the true ones. 
Assumption~{\bf A4} is equivalent to Assumption~{\bf A1} in \cite{TWIST-AOS2079},   Assumption~{\bf A4} in \cite{fan2021alma}
and an equivalent assumption in \cite{MinhTang_arxiv2022}.
Finally, Assumption~{\bf A5}  requires that the sparsity factors are of approximately the same order of magnitude.
The latter guarantees that the  discrepancies between the true and the sample-based   eigenvectors are   
similar  across all layers of the network.
Hypothetically, Assumption~{\bf A5} can be  removed, and one can trace the impact of different scales   $\rhonl$
on the clustering errors. This, however, will make clustering error bounds very complicated, so we leave 
this case for future investigation.

Assumption~{\bf A6} postulates that   matrices $\bP_0\upl$ have enough of non-negligible entries.
Assumption~{\bf A6} naturally holds   in the case of the balanced DIMPLE model~\eqref{eq:model}. Indeed,  
in this case, $\| \bP_0\upl\|^2_F \geq \lowc^2 n^2 K^{-2} \, \|\boB_0\upl\|^2_F$.
Due to Assumption~{\bf A3}, one has $1 = \|\boB_0\upl\|_{\infty} \leq  \|\boB_0\upl\|$ and, therefore,  by
Assumptions~{\bf A1}  and  {\bf A4}
\bes 
\|\boB_0\upl\|_F^2 \geq K_m\, \sig_{K_m}^2 (\boB_0\upl)   \geq C_{\lam}^2\, K_m\, \|\boB_0\upl\|^2  \geq 
C_{\lam}^2\, C_K\, K,
\ees 
which implies  $\| \bP_0\upl\|^2_F \geq C n^2/K$.

Note that Assumption~{\bf A3} implies that $n \to \infty$. In what follows, we assume that $L$ 
can grow at most polynomially with respect to $n$, specifically, that for some constant $\tau_0$
\be \label{eq:nLtau}
L \leq n^{\tau_0}, \quad 0 < \tau_0 < \infty
\ee
Condition \eqref{eq:nLtau} is hardly restrictive. Indeed,    \cite{TWIST-AOS2079} assume  that $L \leq n$, so, in their paper, 
\eqref{eq:nLtau}  holds with $\tau_0=1$. We allow any polynomial growth of $L$ with respect to~$n$.


\subsection{The between-layer clustering error}
\label{sec:error_between}

Evaluation of the between-layer clustering error relies on the 
Tucker decomposition of the tensor with layers  $\bU_{P,l}  (\bU_{P,l})^T$,
$\linL$.
Consider tensor  $\frS \in \RR^{n \times n \times L}$ with layers
\be \label{eq:true_full_tensor}
\frS(:,:,l)= \bU_{P,l}  (\bU_{P,l})^T = \bV\upm (\bV\upm)^T, \quad m = c(l),\ \linL 
\ee
and its clustered version $\calU  \in \RR^{n \times n \times M}$ of the form
\be \label{eq:true_small_tensor}
\calU =   \frS \times_3 [\bC (\bD_{c})^{-1}]^T,  
\ee
where $\bD_{c}$ is defined in \eqref{eq:bC_structure}.
Here,   tensor $\calU$ has layers identical to the set of distinct layers of tensor $\frS$, so that 
$\calU(:,:, m) = \bV\upm (\bV\upm)^T$,   $\minM$.

Recall that, according to \eqref{eq:true_full_tensor} and \eqref{eq:true_small_tensor}, one has
$\frS = \calG \times_3 \bC$. Then, using matrix $\barbZ$ in \eqref{eq:total_basis}, one can rewrite $\frS$ as 
$\frS = \calB \times_1 \barbZ \times_2 \barbZ \times_3 \bC$, 
where $\calB \in \RR^{\barK M \times \barK M \times M}$ is the core tensor with layers
\bes
\calB(:,:,m) = \diag(\bzero_{K_1}, ..., \bzero_{K_{m-1}}, \bI_{K_m}, \bzero_{K_{m+1}}, ..., \bzero_{K_M}) 
\in \{0,1\}^{\barK M \times \barK M}
\ees
Using the SVD in \eqref{eq:total_basis} and the definition of $\bW$ in  \eqref{eq:bC_structure}, we obtain 
\be \label{eq:frS_1}
\frS  = \calF \times_1 \barbU \times_2 \barbU \times_3 \bW,
\quad 
\calF = \barcalR \times_1  \barbD \times_2 \barbD \times_3 \bD_c^{1/2},
\quad 
\barcalR = \calB \times_1 \barbV^T  \times_2 \barbV^T, 
\ee
where $\calF, \barcalR \in \RR^{r \times r \times M}$.
Now, in order to use representation \eqref{eq:frS_1} for analyzing matrix 
$\bTe$ in \eqref{eq:rel1}, note that $\bTe$ is the transpose of mode 3 matricization of $\frS$, i.e., 
$\bTe = \frS_{(3)}^T$.  Using Proposition~1 of \cite{Kolda09tensordecompositions}, obtain
\be \label{eq:bTe_1}
\bTe = (\barbU \otimes \barbU) \bF \bW^T, \quad
\bF = \calF_{(3)}^T \in \RR^{r^2 \times M}.
\ee
Here, by \eqref{eq:bC_structure}  and \eqref {eq:total_basis},  
$\bW = \bC \bD_c^{-1/2} \in \calO_{L,M}$ and  $\barbU  \in \calO_{n,r}$. 
The following statement explores the structure of matrix $\bF$ in \eqref{eq:bTe_1}.

\begin{lem}  \label{lem:bF_struct}
Matrix $\bF$ can be presented  as $\bF = (\barbD  \otimes \barbD) \barbR \bD_c^{1/2}$
where $\barbR = (\barbV \otimes \barbV)^T \bR$ and $\bR =   \calB_{(3)}^T$.
Here, $\rank(\bF) = M$, and, under Assumptions {\bf A1}--{\bf A6}, one has
\be \label{eq:lembF_1}
\sig_{\min}^2(\bF)  = \sig_M^2(\bF) \geq  
\frac{\lowc}{\highc\, \kappa_0^4 M}\, \|\bF\|^2_F \geq 
\frac{\lowc\, C_K\, \barK \, L}{ \highc\, \kappa_0^4\,  M} 
\ee 
\\
\end{lem}

\medskip
 
\noindent
Let the SVD of $\bF$ be of the form $\bF = \bU_F \bLam_F \bV_F$, where $\bU_F \in \calO_{r^2,M}$ and $\bV_F \in \calO_M$.
Then, the SVD of $\bTe$ in \eqref{eq:bTe_1} can be written as
\be \label{eq:SVD_bTe}
\bTe = \bcalV \bLam  \bcalW, \quad \bcalV = (\barbU \otimes \barbU) \bU_F \in \calO_{n^2,M}, \ 
\bcalW  = \bW \bV_F \in \calO_{L,M}, \ \bLam = \bLam_{F}
\ee 
Representation \eqref{eq:SVD_bTe} allows one to bound above the between-layer clustering error.
\\

\begin{thm} \label{th:error_between}
Let Assumptions {\bf A1}--{\bf A6} and \eqref{eq:nLtau} hold. Then, for any $\tau > \tau_0$, there exists a constant 
$C$ that depends only on  $\tau$, $C_K$, $\kappa_0$, $\highc$, $\lowc$, $\highcrho$ 
and $\lowcrho$ in Assumptions {\bf A1}--{\bf A6}, 
such that the between-layer clustering error,  defined in \eqref{eq:err_betw_def}, satisfies
\be \label{eq:error_between}
\PP \lfi R_{BL} \leq 
\frac{C K^2}{ n \rho_n}  \rfi \geq 1 -  L\,  n^{- \tau}  \geq 1 -   n^{-(\tau-\tau_0)}
\ee
\end{thm}


\subsection{The subspace  fitting errors   in groups of layers in the DIMPLE-GDPG model}
\label{sec:error_fitting}

In this section, we provide upper bounds for the divergence between matrices $\bV\upm$ and their estimators
$\hbV\upm$, $\minM$. We measure their discrepancies by $R_{S,max}$ and $R_{S,ave}$ defined in, respectively, 
\eqref{eq:err_subspace_max} and \eqref{eq:err_subspace_ave}.  

\begin{thm} \label{th:error_Vm_est}
Let Assumptions {\bf A1}--{\bf A6} and \eqref{eq:nLtau} hold, and matrices  $\hbV\upm$, $\minM$, 
be obtained using   Algorithm~\ref{alg:subspace_est}. Let
\be \label{eq:nKMcond}
\lim_{n \to \infty} \frac {M K^2}{n \rhon} = 0.
\ee
Then, for any $\tau >0$, there exists a constant $C$  that depends  only on   constants  
in Assumptions {\bf A1}--{\bf A6}, and a constant $C_{\tau, \eps}$ which depends  only on $\tau$ and $\eps$,  
such that  the subspace estimation errors $R_{S,max}$ and $R_{S,ave}$ defined in, respectively, 
\eqref{eq:err_subspace_max} and \eqref{eq:err_subspace_ave}, satisfy
\be \label{eq:error_Smax}
\PP\, \lfi 
R_{S,max} \leq   
C\, \frac{K^{5/2}\, M}{\sqrt{n \, \rhon}} \lkr 1 + \frac{\sqrt{\log n}}{\sqrt{L\, M}} + 
\frac{K\, \sqrt{\log n}}{\sqrt{n \rhon}}  \rkr 
\rfi   \geq 1 - C_{\tau, \eps}\, L\, n^{1 - \tau}
\ee
\be \label{eq:error_Save}
\PP\, \lfi R_{S,ave} \leq   
C\, \frac{K^5 \, M}{n \, \rhon} \lkr 1 + \frac{ \log n}{L} + \frac{K^2\, \log n}{n \rhon} \rkr \rfi
 \geq 1 -  C_{\tau, \eps}\, L\, n^{1 - \tau} 
\ee
\end{thm}

\noindent
Note that, due to condition \eqref{eq:nLtau}, if  $\tau > \tau_0+1$, 
then the upper bounds in \eqref{eq:error_Smax} and \eqref{eq:error_Save}
hold with probability at least $1 - \tilC_{\tau, \eps}\, n^{-(\tau-\tau_0 -1)}$.

\begin{rem} \label{rem:M_eq_1}
{\bf  Subspace estimation error for a homogeneous multilayer GDPG.\ }
{\rm
Consider the case when $M=1$, so that all layers of the network can be embedded into the same invariant subspace.
Since the dominant terms in \eqref{eq:error_Smax} and \eqref{eq:error_Save}
are due to clustering of layers, it follows from the proof of Theorem~\ref{th:error_Vm_est}
in Section~\ref{sec:proof_error_Vm_est}, where  $\|\hbH\upm - \bH\upm\|$ is replaced with $ \Del_{1}\upm$, $m=M=1$, that  
\bes
\left\| \sin\Te \lkr \hbV, \bV  \rkr \right\|_F  
\leq C\  \frac{K^{5/2}\,  \lkv \rhon^{3/2} n^{3/2}  \sqrt{\log n} + \rhon^2\, n  \sqrt{L} \rkv }{n^2\, \rhon^2\,  \sqrt{L}}
= C\,  K^{5/2}\, \lkv \frac{\sqrt{\log n}} {\sqrt{n\, \rhon\,L}} + \frac{1}{n}  \rkv   
\ees
Consequently, one has much smaller subspaces estimation error 
\be \label{eq:M_eq_1}
\PP\, \lfi 
R_{S,max} \leq  C\,  K^{5/2}\, \lkv \frac{\sqrt{\log n}} {\sqrt{n\, \rhon\,L}} + \frac{1}{n}  \rkv 
\rfi   \geq 1 - \tilC_{\tau, \eps}\, L\, n^{1 - \tau}
\ee
} 
\end{rem}


\subsection{The within-layer clustering error}
\label{sec:error_within}

Since the within-layer clustering for each group of layers is  carried out by clustering rows of the matrices 
$\hbV\upm$, the upper bound for $R_{WL}$ defined in \eqref{eq:err_within_ave} can be easily 
obtained as a by-product of Theorem~\ref{th:error_Vm_est}. Specifically, the following statement holds.

\begin{cor} \label{cor:error_within}
Let assumptions of Theorem~\ref{th:error_Vm_est} hold. 
Then, for any $\tau >0$, there exists a constant $C$  that depends  only on   constants  
in Assumptions {\bf A1}--{\bf A6}, and $C_{\tau, \eps}$ which depends  only on $\tau$ and $\eps$,  
such that 
\be \label{eq:within_error_ave}
\PP\, \lfi R_{WL} \leq   
C\, \frac{K^4 \, M}{n \, \rhon} \lkr 1 + \frac{ \log n}{L} + \frac{K^2\, \log n}{n \rhon} \rkr \rfi
 \geq 1 -  C_{\tau, \eps}\, L\, n^{1 - \tau} 
\ee
\end{cor}

\noindent
Note that in the  case of $M=1$, Corollary~\ref{cor:error_within} yields, with high probability, that
\be \label{eq:RWL_M1}
R_{WL} \leq  C\,  K^4\, \lkv \frac{\log n} {n\, \rhon\,L}  + \frac{1}{n^2}   \rkv 
\ee



\section{Simulation study }  
\label{sec:simul_study}
 

\begin{figure}[t]   
\hspace*{-2cm}
\includegraphics[width=18.5cm, height=4.3cm]{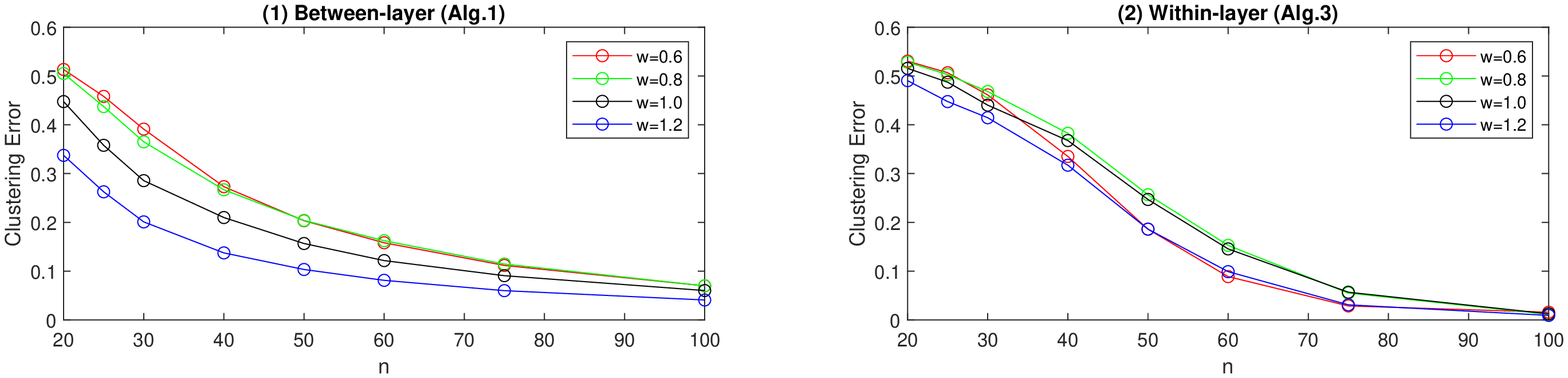}\\
\hspace*{-2cm}
\includegraphics[width=18.5cm, height=4.3cm]{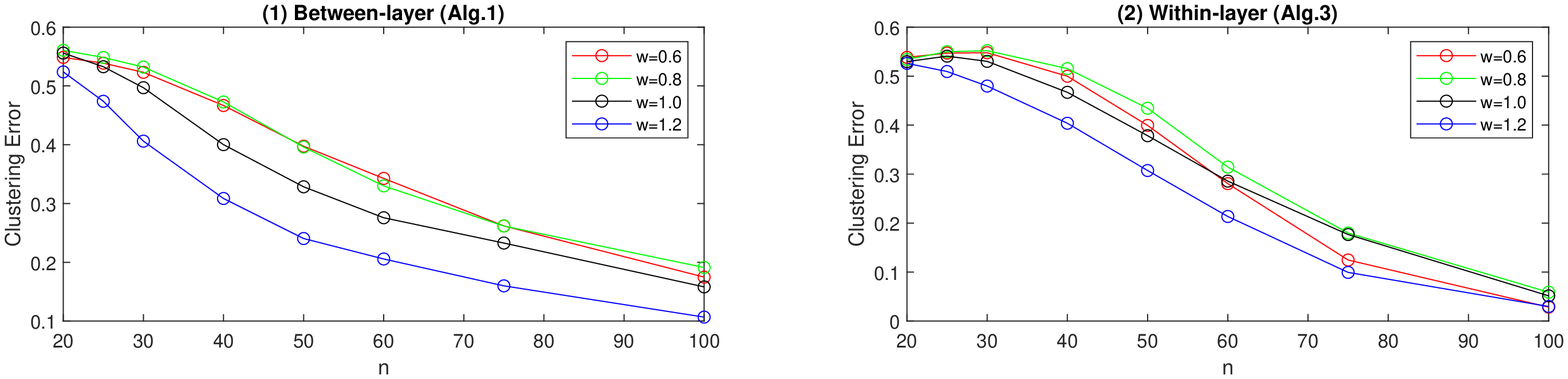} 
\caption{{\small  The between-layer clustering error rates of Algorithm 1 (left) and the within-layer 
error rates of Algorithms~3 (right), averaged over 500 simulation runs,  
for the DIMPLE model with $c = 0, d = 0.8$ (top) and   $c = 0, d = 0.5$ (bottom), $L = 50$ and  
$n = 20, 25, 30, 40, 50, 60, 75, 100$.  The entries of $\bB\upl$, $\linL$, 
are generated  as uniform random numbers  between $c$ and $d$. 
 All the non-diagonal entries of those matrices are subsequently multiplied by $w$.
}}
\label{Sfig_SBM_n}
\end{figure}


In order to study performances of our methodology for various combinations of parameters, 
we carry out a limited simulation study with models generated  from DIMPLE and DIMPLE-GDPG. 
We use Algorithm~1 for finding the groups of layers and Algorithms~2 and 3, respectively, for 
recovering the ambient subspaces in the DIMPLE-GDPG setting, and for finding communities in groups of layers 
for the DIMPLE model. 

To obtain a multilayer network that complies with our assumptions in Section~\ref{sec:assump}, 
we   fix $n$, $L$, $M$, $K$, the  sparsity parameters $c$ and $d$, the assortativity parameter $w$, 
and the Dirichlet parameter $\alpha$ used for generating a DIMPLE-GDPG network. 
We use the multinomial distribution with equal probabilities $1/M$ to assign group memberships to individual networks.

In the case of the DIMPLE model, we generate $K$ communities in each of the groups of layers   
using the multinomial distribution with equal probabilities $1/K$. In this manner, we obtain  community assignment matrices $\bZ{\upm}$,
$\minM$, in each layer $l$ with $c(l)=m$, where $c: [L] \to [M]$ is the layer assignment function. 
Next, we generate the entries of $\bB\upl$, $\linL$, as uniform random numbers  between $c$ and $d$, 
and then multiply all the non-diagonal entries of those matrices by $w$.  In this manner, if 
 $w<1$ is small, then the network is strongly assortative, i.e., there is a higher  probability for nodes in the same community to connect.
If $w>1$ is large,  then the network is disassortative, i.e., the probability of connection for nodes  
in different communities is higher than for nodes in the same community. Finally, since entries 
of matrices $\bB\upl$ are generated at random, when $w$ is close to one, the networks in all layers
are neither assortative or disassortative. 
After the community assignment matrices $\bZ{\upm}$ and the block probability matrices $\bB\upl$ have been obtained, 
we construct the probability tensor $\calP$ with layers   $\calP (:,:,l) = \bZ{\upm}\bB\upl(\bZ{\upm})^T$,
where $m = c(l)$, $\linL$.


\begin{figure}[t]    
\hspace*{-2cm}  
\includegraphics[width=18.5cm, height=4.3cm]{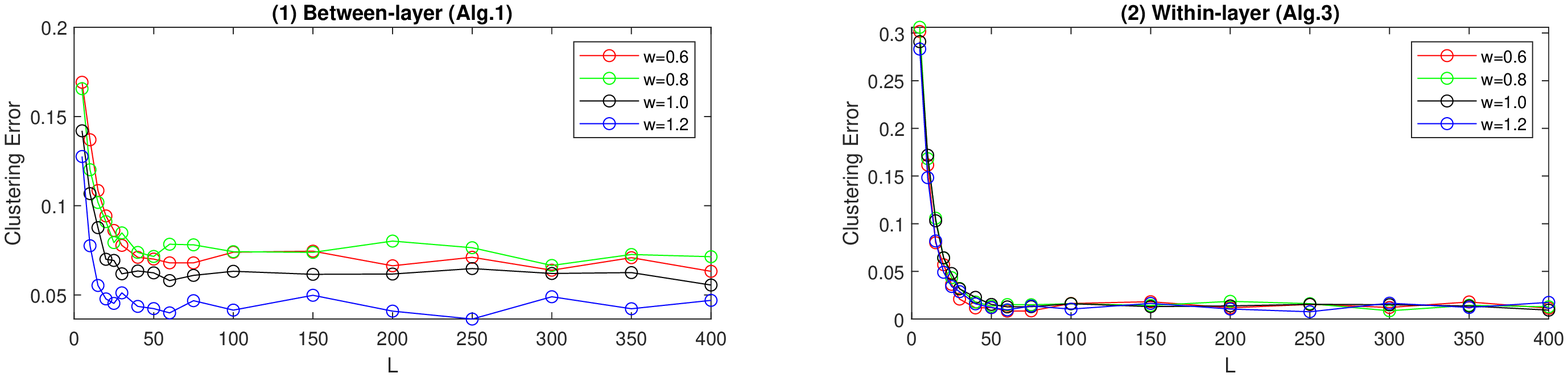}\\
 \hspace*{-2cm}  
\includegraphics[width=18.5cm, height=4.3cm]{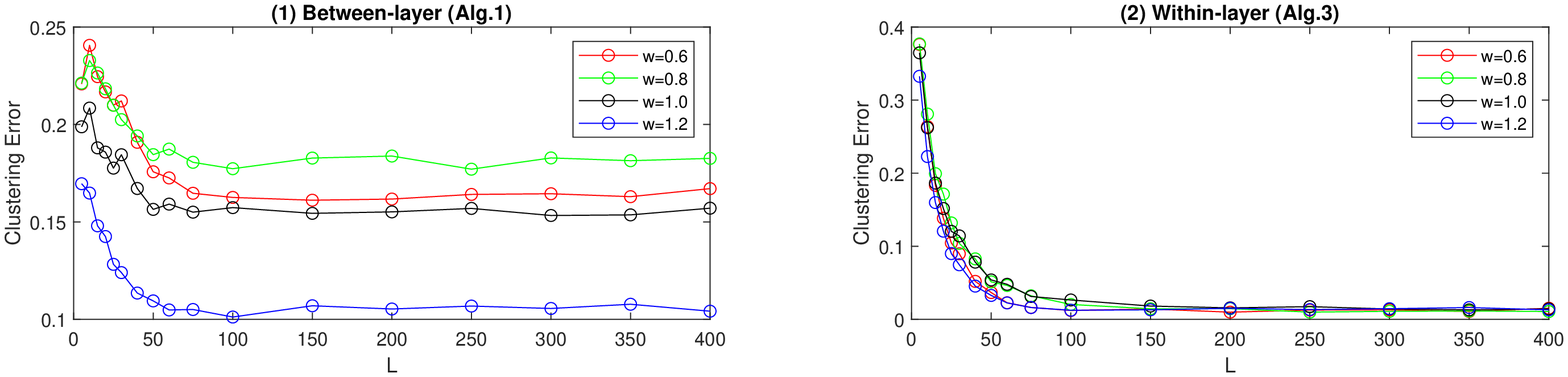}
\caption{{\small  The between-layer clustering error rates of Algorithm 1 (left) and the within-layer 
error rates of Algorithms~3 (right), averaged over 500 simulation runs,  
for the DIMPLE model with $c = 0, d = 0.8$ (top) and   $c = 0, d = 0.5$ (bottom), $n = 100$ and  
$L = 5, 10, 15, 20, 25, 30, 40, 50, 60, 75, 100, 150, 200, 250, 300, 350, 400$.  
The entries of $\bB\upl$, $\linL$, are generated  as uniform random numbers  between $c$ and $d$. 
 All the non-diagonal entries of those matrices are subsequently multiplied by $w$.
}}
\label{Sfig_SBM_L}
\end{figure}


In the case of the DIMPLE-GDPG setting, we obtain matrices $\bX\upm \in [0,1]^{n \times K}$, $\minM$,  with 
independent rows, generated using the Dirichlet distribution with parameter $\alpha$. We obtain matrices 
$\bB\upl$,   in exactly the same manner as in the case of the DIMPLE model and construct 
$\calP$ with layers   $\calP (:,:,l) = \bX{\upm}\bB\upl(\bX{\upm})^T$, where $m = c(l)$, $\linL$.
In this case, the matrices $\bV\upm$ are obtained from  the SVD   $\bX\upm = \bV\upm \bLam_X\upm \bW_X\upm$ 
of  $\bX\upm$. Matrices $\bQ\upl$ are defined as  $\bQ\upl = \bLam_X\upm \bW_X\upm \bB\upl (\bW_X\upm)^T \bLam_X\upm$ 
in \eqref{eq:DIMPLE_GDPG}, $\linL$.

After the probability tensor $\calP$ is generated, the layers $\bA\upl$ of the adjacency tensor $\calA$
are obtained as symmetric matrices with zero diagonals and independent Bernoulli entries 
$\bA\upl (i,j)$ for $1 \leq i <j \leq n$. Subsequently, we use Algorithm~1 for finding the groups
of layers for both models, followed by Algorithm~2 for estimating matrices $\bV\upm$ in the case of 
the DIMPLE-GDPG network, or Algorithm~3 for clustering nodes in each group of layers of the network into communities 
for the DIMPLE model. In both cases, we have two sets of simulations, one with fixed $L$ and varying $n$, 
another with the fixed $n$ and varying $L$. In all simulations, we set $M=3$ and $K_m=3$ for $m=1,2,3$,  and study two sparsity scenarios,
$c = 0$, $d = 0.8$ or $c = 0$, $d = 0.5$,  with four values of assortativity parameter $w = 0.6, 0.8, 1.0$ and $1.2$.  
In all simulations, we set $\alpha = 0.1$. 
We report the average between-layer clustering errors $R_{BL}$ defined in \eqref{eq:err_betw_def}, and also
the average within-layer clustering error $R_{WL}$ defined in \eqref{eq:err_within_ave} in the case of the DIMPLE setting
and the average $\sinTe$ distance $R_{S,ave}$ defined in \eqref{eq:err_subspace_ave}  between the true and the estimated subspaces 
in the case of the DIMPLE-GDPG network. We first present simulations  results for the DIMPLE model followed 
by the study of the DIMPLE-GDPG  model.


\begin{figure}[t]  
\hspace*{-2cm}
\includegraphics[width=18.5cm, height=4.3cm]{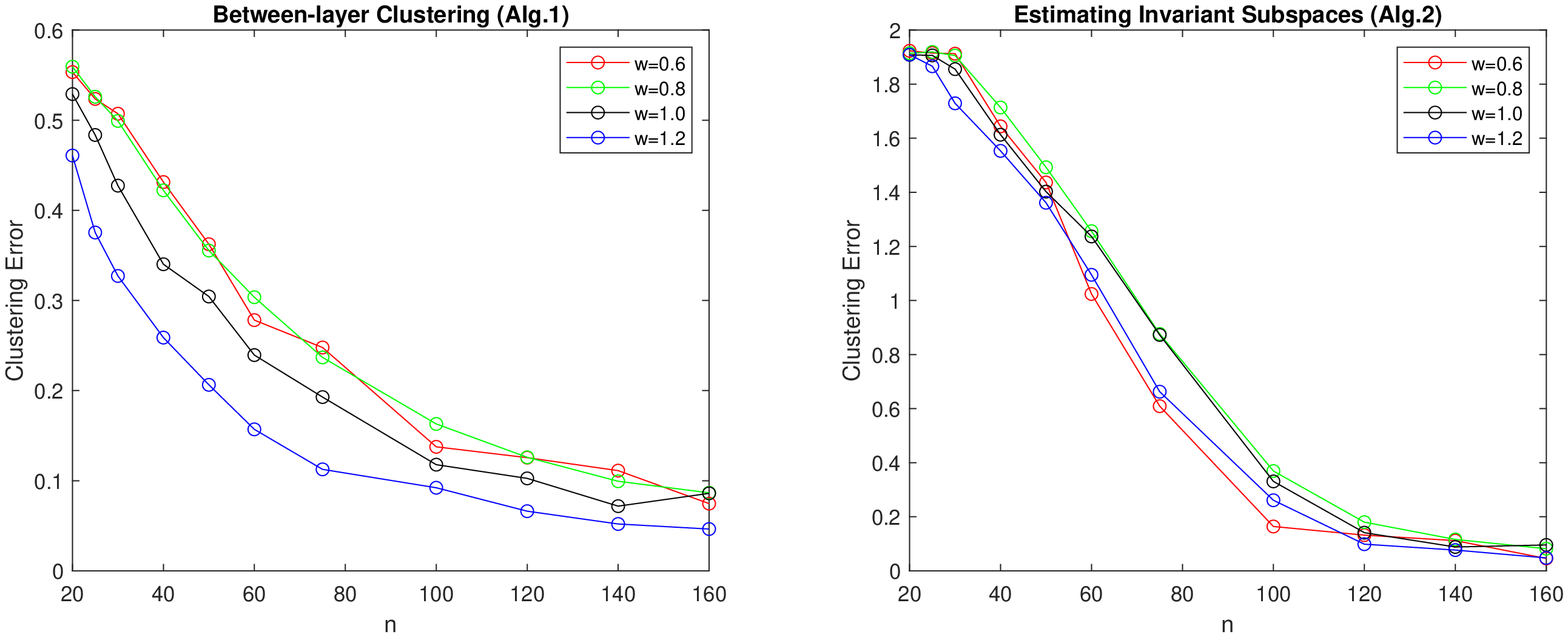}
\hspace*{-2cm}
\includegraphics[width=18.5cm, height=4.3cm]{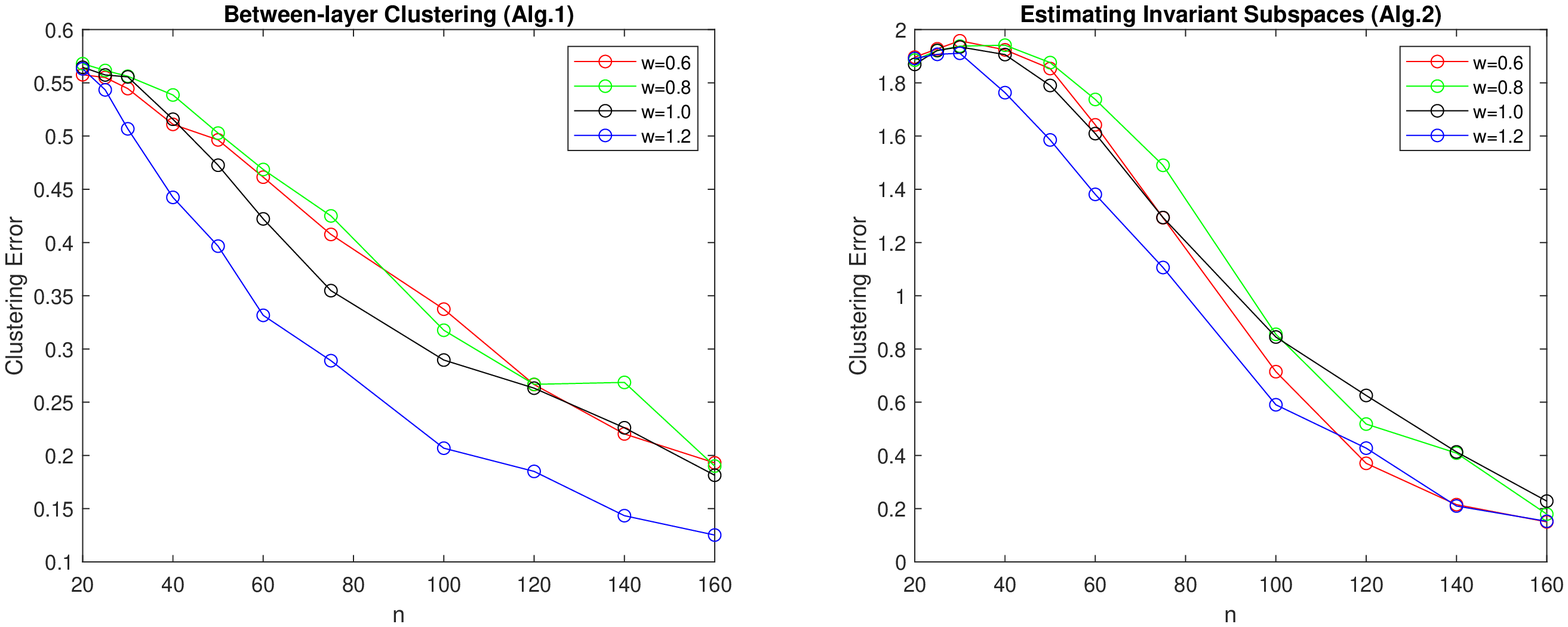}
\caption{{\small  The between-layer clustering error rates of Algorithm 1 (left) and 
the   $\sinTe$ distances $R_{S,ave}$ of Algorithms~2 (right), averaged over 100 simulation runs,  
for the DIMPLE-GDPG model with $\alpha = 0.1$, $c = 0, d = 0.8$ (top) and   $c = 0, d = 0.5$ (bottom), $L = 50$ and  
$n = 20, 25, 30, 40, 50, 60, 75, 100, 120, 140, 160$.  The entries of $\bB\upl$, $\linL$, 
are generated  as uniform random numbers  between $c$ and $d$. 
 All the non-diagonal entries of those matrices are subsequently multiplied by $w$.
}}
\label{Sfig_GDPG_n}
\end{figure}



\begin{figure}[t]  
\hspace*{-2cm}
\includegraphics[width=18.5cm, height=4.3cm]{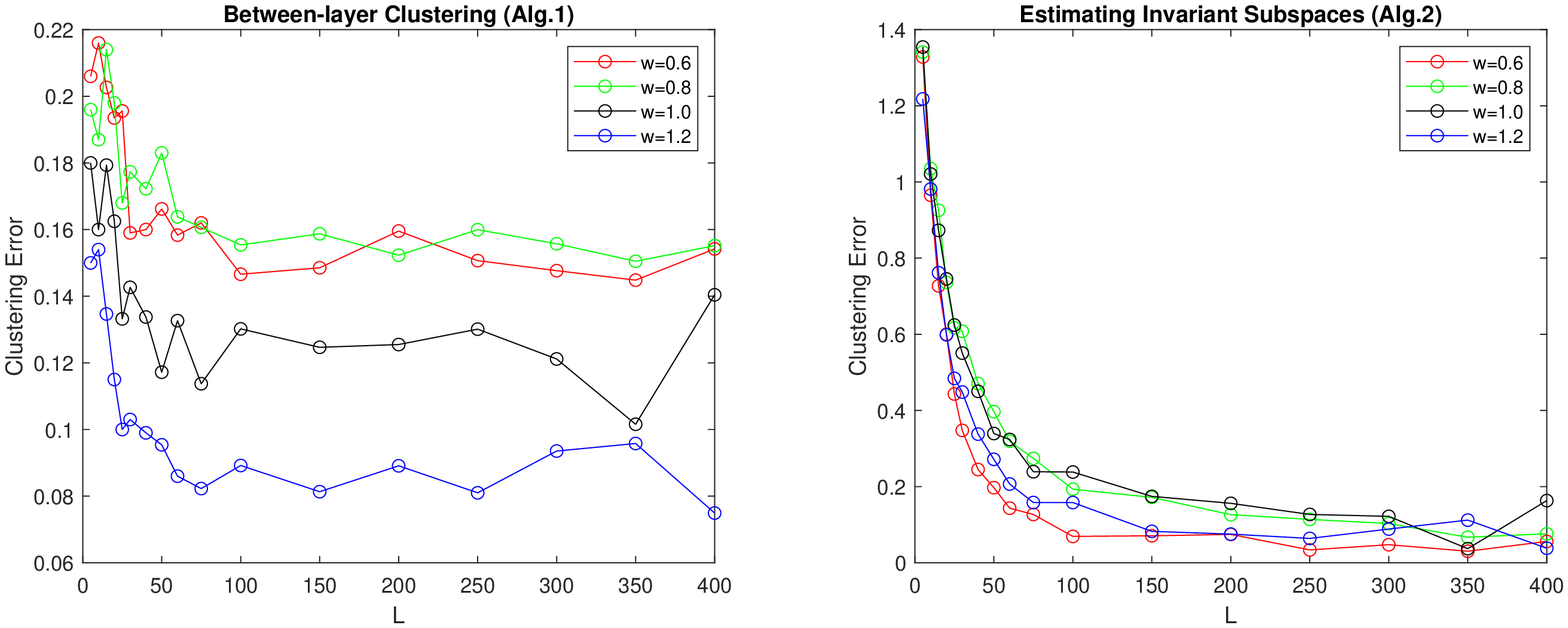}
\hspace*{-2cm}
\includegraphics[width=18.5cm, height=4.3cm]{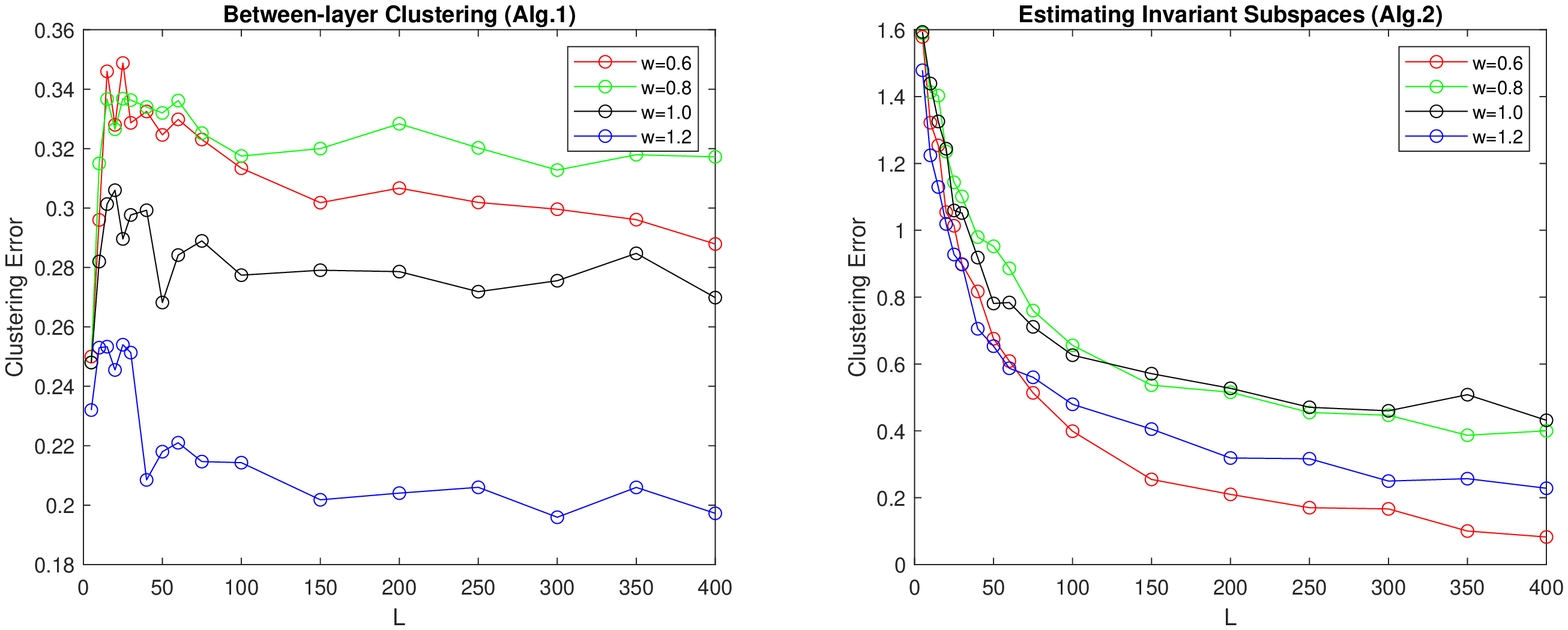}
\caption{{\small  The between-layer clustering error rates of Algorithm 1 (left) and 
the   $\sinTe$ distances $R_{S,ave}$ of Algorithms~2 (right), averaged over 100 simulation runs,  
for the DIMPLE-GDPG model with $\alpha = 0.1$,  $c = 0, d = 0.8$ (top) and   $c = 0, d = 0.5$ (bottom), $n = 100$ and  
$L = 5, 10, 15, 20, 25, 30, 40, 50, 60, 75, 100, 150, 200, 250, 300, 350, 400$.  The entries of $\bB\upl$, $\linL$, 
are generated  as uniform random numbers  between $c$ and $d$. 
 All the non-diagonal entries of those matrices are subsequently multiplied by $w$.
}}
\label{Sfig_GDPG_L}
\end{figure}


Simulations results for the DIMPLE and DIMPLE-GDPG models are summarized in Figures~\ref{Sfig_SBM_n}--\ref{Sfig_SBM_L}
and  Figures~\ref{Sfig_GDPG_n}--\ref{Sfig_GDPG_L}, respectively. Note that, while the between-layer clustering errors 
(left panels in Figures~\ref{Sfig_SBM_n}--\ref{Sfig_GDPG_L}), as well as the within-layer clustering errors 
(right panels in Figures~\ref{Sfig_SBM_n}--\ref{Sfig_SBM_L}) are between 0 and 1,   
the average errors of estimation of subspaces $R_{S,ave}$ defined in \eqref{eq:err_subspace_ave}
(right panels in Figures~\ref{Sfig_GDPG_n}--\ref{Sfig_GDPG_L}) lie between 0 and $K$, so they are on a different scale.

As it is expected, both estimation and clustering are harder 
when a network is more sparse, therefore, all errors are smaller when $d=0.8$ (top panels) than when $d=0.5$ (bottom).
Figures~\ref{Sfig_SBM_n}--\ref{Sfig_GDPG_L} show that the value of the assortativity parameter does not play a significant role 
in the between-layer clustering. Indeed, as the left panels in all figures show, the smallest between-layer clustering
errors occur for $w=1.2$ followed by $w=1.0$. The latter confirms that the difficulty of the between-layer clustering
is predominantly controlled by the sparsity of the network. 
The results are somewhat different for the community detection errors  and the subspace 
estimation errors in, respectively,  the DIMPLE and the DIMPLE-GDPG models.
Indeed, as the   right panels in Figures~\ref{Sfig_SBM_n}--\ref{Sfig_GDPG_L}  show, 
the smallest errors occur in the more assortative/disassortative  models with 
$w=0.6$ and $w=1.2$.

One can see  from Figures~\ref{Sfig_SBM_n} and \ref{Sfig_GDPG_n}    that, when $n$ grows, all  errors decrease.
The influence of $L$ on the error rates is more complex. As Theorem~\ref{th:error_between} 
implies, the between-layer clustering errors are of the order $(n \rhon)^{-1}$ 
for fixed values of  $M$ and $K$. This agrees with the left panels in Figures~\ref{Sfig_SBM_L} and \ref{Sfig_GDPG_L}
where curves exhibit constant behavior for when $L$  grows (small fluctuations are just due to  random errors).
For the right panels in Figures~\ref{Sfig_SBM_L} and \ref{Sfig_GDPG_L} this,  however, happens only when $L$ 
is relatively large.

The explanation for such behavior lies in the fact that the between-layer clustering error (corresponding to the left panels
in Figures~\ref{Sfig_SBM_L} and \ref{Sfig_GDPG_L}) is of the order $K^2\, (n \rhon)^{-1}$
and is independent of $L$. On the other hand, for fixed $K$ and $M$, the errors $R_{WL}$ and $R_{S,ave}$ (corresponding to the right panels
in, respectively, Figures~\ref{Sfig_SBM_L} and \ref{Sfig_GDPG_L}) are of the order 
$(n \, \rhon)^{-1} + \log n\, (n \, \rhon\, L)^{-1}$.  While $L$ is small the second term is dominant but, as $L$ grows.
the first term becomes dominant and the errors stop declining as $L$ grows.


\section{Application to the Real World Data}
\label{sec:real_data}

In this section, we consider applications of the DIMPLE and the DIMPLE-GDPG models to real-life data,
and its comparison with the MMLSBM. Note that the between-layer clustering is carried out by Algorithm~\ref{alg:between}
for both the DIMPLE and the DIMPLE-GDPG models, so one can decide which of the models to use later in the analysis.

In our examples, the DIMPLE model with its SBM-imposed structures provided better descriptions of the 
organization of  layers in each group than its GDPG-based  DIMPLE-GDPG counterpart.
Furthermore, we compared our between layer clustering partitions with the ones obtained on the basis of the MMLSBM
setting.


\subsection{Worldwide Food Trading Network Data}
\label{sec:food}

In this subsection, we consider applying  our clustering algorithms to the Worldwide Food Trading Networks data collected 
by the Food and Agriculture Organization of the United Nations.
The data have been described  in   \cite{doi:10.1038/ncomms7864}, and it is available at \url{https://www.fao.org/faostat/en/#data/TM}. 
The data includes export/import  trading volumes among  245  countries for more than  300  food items.
These data can be modeled as   a multiplex network,  in which layers represent different products, nodes are countries,
and edges at each layer represent trading relationships of a specific food product among countries. 
A part of the  data set  was analyzed in  \cite{TWIST-AOS2079}  and \cite{fan2021alma}.

Similarly to  \cite{TWIST-AOS2079}  and \cite{fan2021alma}, we used data for the year 2010.
We start with pre-processing the data by adding the export and import volumes for each pair of countries in each layer of the network,
to produce undirected networks that  fit in our model. To avoid sparsity, we select 104 countries, whose total trading volumes are  higher 
than the median among all countries. 
We choose 58 meat/dairy  and fruit/vegetable  items and constructed a network with 104 nodes and 58 layers.

While pre-processing the data, we  observe that global trading patterns are different for the meat/dairy and the fruit/vegetable 
groups. Specifically,  the trading volumes in meat/dairy group are much smaller than the trading volumes in the 
fruit/vegetable  group. For this reason, we choose the thresholds that keep similar sparsity levels for the adjacency matrices.
In particular, we set threshold to be equal to 1 unit for the meat/dairy  group  and 300 units  for the fruit/vegetable
group, and draw an edge between  two nodes (countries) if  the total trading volume between them is at or above the threshold.

\begin{figure}[htpb]   
\hspace*{1.1cm}  
\includegraphics[width=12.5cm, height=8.3cm]{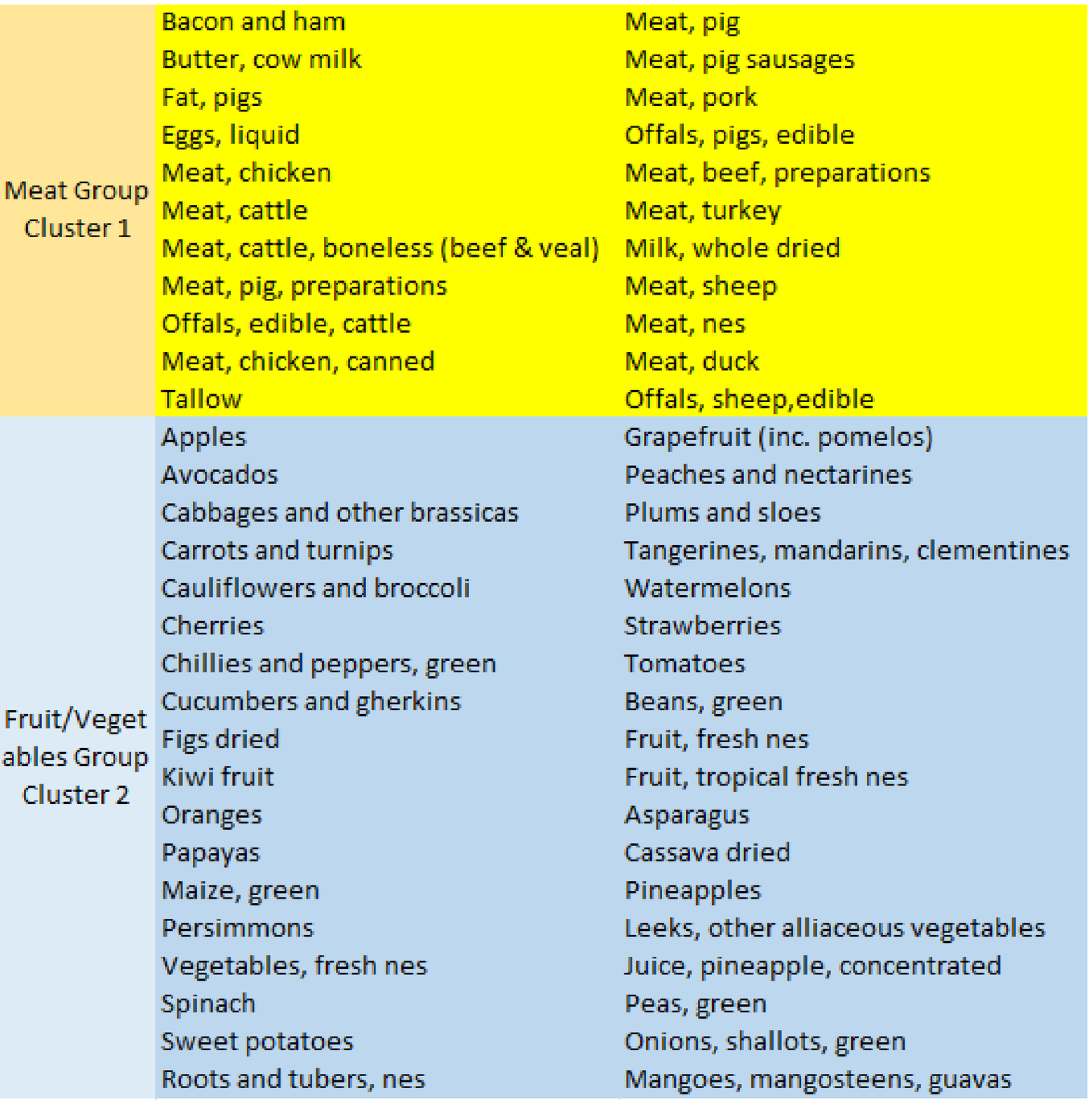}
\caption{ Results of clustering of food  networks  layers into $M=2$ clusters by Algorithm~1 in the paper
}\label{Food_fig1}
\end{figure}

We scramble  the 58 layers and apply Algorithm~1 for the  between-layer clustering.
 Since the food items consist of a meat/dairy and a fruit/vegetable group, we set $M=2$. 
Due to the fact that there are five  food regions (continents)  in the world, Asia, America, Europe, Africa and Australia, 
we  start  with the number of communities in each layer to be $K=5$. However, the latter  leads  to an unbalanced community structure, 
specifically, two communities that consists of only one  country. For this reason, after experimenting, we set $K=3$.
Results of the between-layer clustering are presented in Figure~\ref{Food_fig1}.  
As it is evident from Figure~\ref{Food_fig1}, Algorithm~1 separates the food items into the meat/dairy and the fruit/vegetable 
groups.

\begin{figure}[htpb]   
\hspace*{1.1cm}  
\includegraphics[width=12.5cm, height=8.3cm]{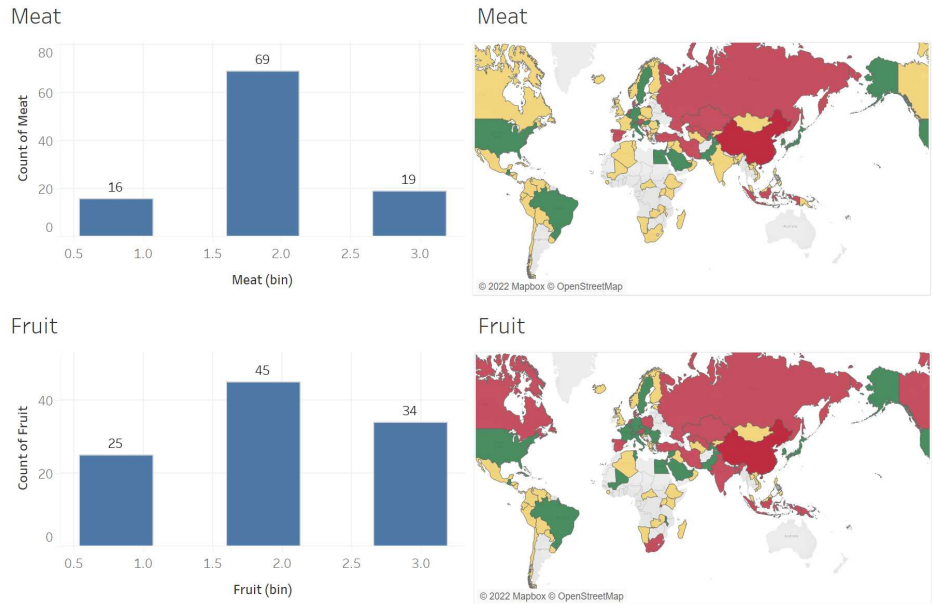}
\caption{{\small  Trading communities for the meat/dairy (top)  and the fruit/vegetable (bottom)
groups. Left panels: community sizes; right panels: community memberships 
}}\label{Food_fig2}
\end{figure}

\begin{figure}[htpb]
\centering 
\includegraphics[width=12.5cm, height=8.3cm]{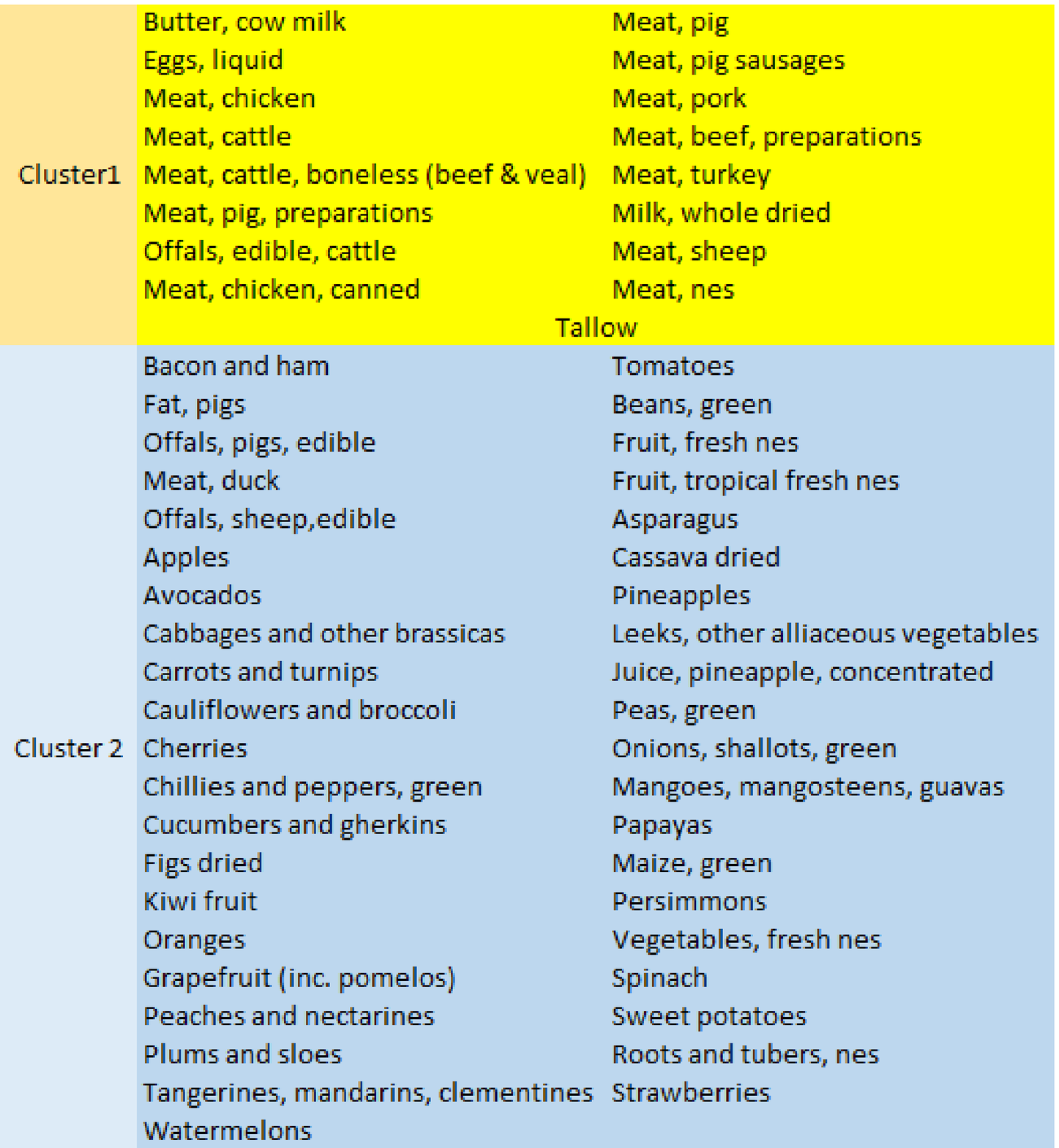}
\caption{ Results of clustering of food  networks  layers into $M=2$ clusters by ALMA algorithm of \cite{fan2021alma}
}\label{Food_fig3}
\end{figure}

Furthermore, we investigate the communities of countries that form trade clusters in each of the two layers.
We use Algorithm~3 in the  paper, and exhibit results of the within-layer clustering   in Figure~\ref{Food_fig2}.
The left panels in Figure~\ref{Food_fig2} show the number of nodes (countries)  in communities 1,2 and 3 in the meat/dairy and the 
fruit/vegetable group, respectively. The right panels in  Figure~\ref{Food_fig2} project those countries onto the world map.
Here, the red color is used for community 1,  the yellow color  for community 2 and  the green color for  for community 3. 
Since we only select 104 countries to be a part of the network, some regions in the map are colored grey.

In order to justify application of the DIMPLE model, we also carry out data analysis assuming 
that data were generated using the MMLSBM. 
Specifically, we applied ALMA algorithm of \cite{fan2021alma} for the layer clustering with the same parameters $M=2$ and $K=3$. 
Results are  presented in Figure~\ref{Food_fig3}. It is easy to notice that  ALMA algorithm places  some of the meat/dairy items 
into the  fruit/vegetable group. We believe that this is due to the fact that MMLSBM is sensitive to the probabilities 
of connections rather than connection patterns.


\subsection{Global Flights Network Data}
\label{sec: flight}


\begin{table}[t]
\centering
\footnotesize
\begin{tabular}{|c|c||c|c|}
\hline
    \multicolumn{4}{|c|}{Airlines Groups under the DIMPLE-GDPG  Model} \\
  \hline
 \multicolumn{2}{|c|}{{\bf Group 1}}& \multicolumn{2}{|c|}{{\bf Group 2}} \\
 \hline
China &	Hainan Airlines             &   New Zealand  &	Air New Zealand \\
\hline
China &	Air China    & Republic of Korea  &	Korean Air \\
\hline
China &	Sichuan Airlines  & Singapore  &	Singapore Airlines\\
\hline
China &	Shenzhen Airlines  & Australia  &	Qantas\\
\hline
China &	China Southern Airlines  & Vietnam  &	Vietnam Airlines\\
\hline
China &	Shandong Airlines   &  India  &	Air India Limited \\
\hline
China &	China Eastern Airlines  & India  &	IndiGo Airlines \\
\hline
China &	Xiamen Airlines  & Australia  &	Virgin Australia \\
\hline
Japan &	Japan Air System & South Africa  &	South African Airways\\
\hline
 \multicolumn{2}{|c||}{{\bf Group 3} } & Indonesia  &	Garuda Indonesia\\
\hline
Germany &	Lufthansa  & Republic of Korea  &	Asiana Airlines\\
\hline
Russia &	Ural Airlines &  Malaysia   &  Malaysia Airlines\\
\hline
Switzerland &	Swiss International Air Lines & India	&  Jet Airways\\
\hline
Morocco &	Royal Air Maroc  & Japan   &  Japan Airlines\\
\hline
Norway &	Norwegian Air Shuttle  &    Japan   &  All Nippon Airways\\
\hline
Ireland &	Ryanair   & Qatar   &  Qatar Airways\\
\hline
Turkey &	Turkish Airlines  &  Saudi Arabia  &	Saudi Arabian Airlines\\
\hline
Greece &	Aegean Airlines  &   United Arab Emirates   &  Emirates \\
\hline
Algeria &	Air Algerie  & United Arab Emirates   &  Etihad Airways\\
\hline
Ethiopia &	Ethiopian Airlines  &  \multicolumn{2}{|c|}{{\bf Group 4}} \\
\hline
United Kingdom &	Jet2.com & United States   &  JetBlue Airways\\
\hline
United Kingdom &	Flybe  & United States   &  US Airways\\
\hline
Russia &	Transaero Airlines  & United States   &  Alaska Airlines\\
\hline
Germany &	Condor Flugdienst & United States   &  Southwest Airlines\\
\hline
Germany &	TUIfly  & United States   &  Delta Air Lines\\
\hline
Sweden &	Scandinavian Airlines & United States   &  AirTran Airways\\
\hline
Portugal &	TAP Portugal  & United States   &  Spirit Airlines\\
\hline
France &	Transavia France  & United States   &  United Airlines\\
\hline
United Kingdom &	British Airways  & United States   &  American Airlines\\
\hline
Russia &	S7 Airlines  & United States   &  Frontier Airlines\\
\hline
Ireland &	Aer Lingus  & Canada   &  Air Canada\\
\hline
Germany &	Germanwings  & Canada   &  WestJet\\
\hline
Egypt &	Egyptair   & Mexico   &  AeroMexico\\
\hline
Austria &	Austrian Airlines  & Chile   &  LAN Airlines\\
\hline
Spain &	Iberia Airlines  & Brazil   &  TAM Brazilian Airlines\\
\hline
Germany &	Air Berlin  & South America   &  Avianca\\
\hline
Italy &	Alitalia  & Netherlands   &  KLM Royal Dutch Airlines\\
\hline
Hungary &	Wizz Air  &   France   &  Air France\\
\hline
Finland &	Finnair  & & \\
\hline
Russia &	Aeroflot  & & \\
\hline
France  &	Air Bourbon & & \\
\hline
Netherlands &	Transavia Holland& & \\
\hline
United Kingdom & easyJet & & \\
\hline
\end{tabular}
\caption{\footnotesize Airlines Groups obtained using Algorithm~\ref{alg:between} with $K=3$ and $M=4$
}
\label{table1}
\end{table}


In this subsection, we   applied   our clustering algorithms to the Global Flights Network data collected by the OpenFlights.
As of June 2014, the OpenFlights Database contains 67663 routes between 3321 airports on 548 airlines spanning the globe.
It is available at  
\url{https://openflights.org/data.html#airport}. 

These data can be modeled as a multiplex network,  in which layers represent different airlines, nodes are airports where airlines depart and land, 
and edges at each layer represent existing routes of a specific airline company between two  airports. 
To avoid sparsity, we selected 224 airports, where over 150 airline companies have rights to depart and land in. 
Furthermore, we chose 81 airlines that have at least 240 routes between those airports, constructing a network with 224 nodes and 81 layers.

We scrambled   the 81 layers and applied Algorithm~\ref{alg:between}  for the  between-layer clustering. 
After experimenting with various values of $M$ and $K$, we partitioned the airlines into  $M=4$ groups, and
used the ambient dimension $K=3$ for each of the groups. 
 Results of the between-layer clustering are presented in Table~\ref{table1}.


\begin{table}[t]
\centering
\footnotesize
\begin{tabular}{|c|c||c|c|}
\hline
    \multicolumn{4}{|c|}{Airlines Groups under the MMLSBM   } \\
  \hline
 \multicolumn{2}{|c|}{{\bf Group 1}} & \multicolumn{2}{|c|}{{\bf Group 2}} \\
 \hline
Japan  & Japan Air System  & China  & Hainan Airlines\\
 \hline
China  & Sichuan Airlines  & China  & Air China\\
 \hline
China  & Shandong Airlines  & China  & Shenzhen Airlines\\
 \hline
 China  & Xiamen Airlines  & China  & China Southern Airlines\\
 \hline
Republic of Korea  & Korean Air  & China  & China Eastern Airlines\\
 \hline
Singapore  & Singapore Airlines  &  & \\ 
 \hline
Vietnam  & Vietnam Airlines  &  \multicolumn{2}{|c|}{{\bf Group 3}}\\
 \hline
India  & Air India Limited &  France  & Air France \\
 \hline
 United States &	US Airways & United States  & Delta Air Lines\\
 \hline
Australia  & Qantas  & United States  & AirTran Airways\\
 \hline
Mexico  & AeroMexico & United States  & Southwest Airlines \\
\hline
India  & IndiGo Airlines  & United States  & American Airlines\\
 \hline
South Africa  & South African Airways  &  Netherlands  & KLM Royal Dutch Airlines\\
 \hline
Indonesia  & Garuda Indonesia &  Italy  & Alitalia\\
 \hline
Republic of Korea  & Asiana Airlines & & \\
 \hline
Saudi Arabia  & Saudi Arabian Airlines  & \multicolumn{2}{|c|}{{\bf Group 4}}\\
 \hline
Hong Kong  	& Cathay Pacific & France  & Transavia France\\
 \hline
South America  & Avianca & France  & Air Bourbon\\
 \hline
Japan  & Japan Airlines & United Kingdom  & Jet2.com \\
\hline
Qatar  & Qatar Airways &   United Kingdom  & easyJet\\
\hline
Australia  & Virgin Australia &  Ireland  & Ryanair\\
\hline
Japan  & All Nippon Airways &   & \\
\hline
Malaysia  & Malaysia Airlines &  \multicolumn{2}{|c|}{{\bf Group 1: Continuation}}\\
\hline
India  & Jet Airways & Canada  & WestJet \\
\hline
United Arab Emirates  & Etihad Airways & United Arab Emirates  & Emirates \\
\hline
Germany  & Lufthansa & Russia  & Ural Airlines \\
\hline
Turkey  & Pegasus Airlines & Morocco  & Royal Air Maroc \\
\hline
Switzerland  & Swiss International Airlines & Turkey  & Turkish Airlines \\
\hline
Norway  & Norwegian Air Shuttle & Ethiopia  & Ethiopian Airlines \\
\hline
Greece  & Aegean Airlines & Algeria  & Air Algerie \\
\hline
United Kingdom  & Flybe & Germany  & Condor Flugdienst \\
\hline
Germany  & TUIfly & Sweden  & Scandinavian Airlines \\
\hline
Portugal  & TAP Portugal & United Kingdom  & British Airways \\
\hline
Russia  & S7 Airlines & Austria  & Austrian Airlines \\
\hline
Ireland  & Aer Lingus & Spain  & Iberia Airlines \\
\hline
Germany  & Germanwings & Russia  & Aeroflot  \\
\hline
Egypt  & Egyptair & Germany & Air Berlin \\
\hline
Hungary  & Wizz Air & Russia  & Transaero Airlines\\
\hline
Finland  & Finnair & United States  & Alaska Airlines \\
\hline
Netherlands  & Transavia Holland & Brazil  & TAM Brazilian Airlines \\
\hline
United States  & JetBlue Airways & United States  & Spirit Airlines \\
\hline
Chile  & LAN Airlines & Canada  & Air Canada \\
\hline
New Zealand  & Air New Zealand &  United States  & Frontier Airlines \\
\hline
United States  & United Airlines & &\\
\hline
\end{tabular}
\caption{\footnotesize Airlines Groups obtained using ALMA algorithm of \cite{fan2021alma}  with $K=3$ and $M=4$}
\label{table2}
\end{table}

\begin{figure}[t] 
\centering
\includegraphics[width = 14cm, height = 7cm]{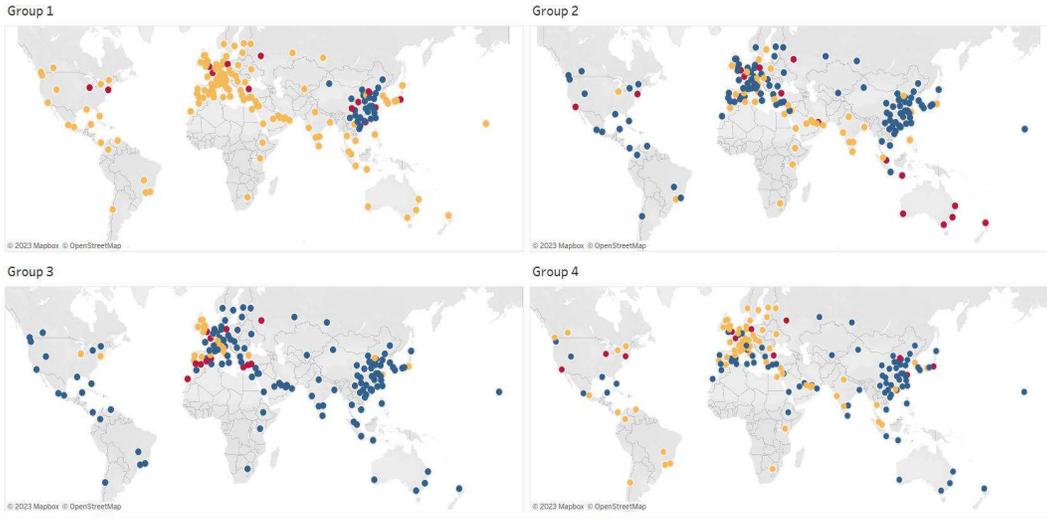}
\caption{\footnotesize Communities for the four airlines groups. Group 1: airlines originated in China. 
Group 2: airlines originated in Asia, Australia,New Zealand, and  Gulf States. Group~3:   airlines originated in  Europe
and North Africa. Group~4:   airlines originated in North or South America. }
\label{fig:air_commun}
\end{figure}

We also partitioned airports in each of the groups of airlines into communities. 
Results are presented  in Figure~\ref{fig:air_commun}.   
 
It is easy to see that in Table~\ref{table1},  the airlines are naturally grouped by geographical areas from where the flights are originated.
Group~1 is constituted by Chinese airline and one Japanese airline which has   flights predominantly in Far East. 
Group~2 consists of airlines that belong to countries in Asia, such as India, Japan, South Korea and Vietnam, 
Australia and New Zealand, and few big airlines in   Gulf States (Saudi Arabia, United Arab Emirates, Qatar)
that have a large number of flights to both Asia and Australia. Group~3 is formed by airlines originated from Europe
and North Africa while Group~4 is comprised of airlines that fly in or from North or South America. 
Not surprisingly, this group includes two big European airlines, KLM and Air France, since those airlines are 
members of the SkyTeam alliance and share many flights originated in USA with Delta airlines.

We also analyzed the airline data under the assumption that they follow the MMLSBM. 
To this end, we applied ALMA algorithm of \cite{fan2021alma} for the layer clustering, with the same parameters $M = 4$ and $K = 3$. 
Results are presented in Table~\ref{table2}.  It is easy to see that while the DIMPLE  model ensures a logical 
geography-based partition of the airlines, the MMLSBM does not. Indeed, the MMLSBM lumps almost all airlines into Group~1,
placing few Chinese airlines into Group~2,  few United States owned airlines together with Air France, Alitalia and KLM into Group~3, 
and Ryanair (Ireland), Transavia and Air Bourbon  (France),   easyJet and Jet2.com (United Kingdom) into Group~4. 
On the contrary, Algorithm~\ref{alg:between} associated with the DIMPLE  model delivers four balanced (similar in size) groups.
 This is due to the fact that MMLSBM groups airlines by the volume of operation rather than the structure of roots.




\section{Discussion}
\label{sec:discussion}

In this paper, we introduce the GDPG-equipped  DIMPLE-GDPG multiplex network model where layers can be partitioned into groups 
with similar ambient subspace structures while the matrices of connections probabilities can be all different. 
In the common case when each layer follows the SBM, the latter reduces to the DIMPLE model, where   community affiliations
are common for each group of layers while the matrices of block  connection probabilities   vary from one layer to another. 
The DIMPLE-GDPG model  generalizes the COmmon Subspace Independent Edge (COSIE) random graph model 
of \cite{JMLR:v22:19-558} and  \cite{MinhTang_arxiv2022}, while the DIMPLE model generalizes a multitude of 
the SBM-equipped multiplex network settings. Specifically, it includes, as its particular cases, 
the Mixture MultiLayer Stochastic Block Model  (MMLSBM) of  \cite{Stanley_2016},  \cite{TWIST-AOS2079} and \cite{fan2021alma},
and the multitude of papers that assume that communities persist in all layers of the network
(see, e.g., \cite{bhattacharyya2020general}, \cite{lei2021biasadjusted}, \cite{10.1093/biomet/asz068},
 \cite{paul2016}, \cite{paul2020}).

Our real data examples in Section~\ref{sec:real_data} show that our models deliver more 
understandable description of data than the MMLSBM, due to the flexibility of the DIMPLE and DIMPLE-GDPG models.

If $M=1$, the DIMPLE-GDPG reduces to COSIE model, and we believe that our paper provides some improvements due to 
employment of a different algorithm for the   matrix $\bV$ estimation. 
Indeed,  \cite{JMLR:v22:19-558} showed that 
\be \label{eq:M1_COSIE}
\EE\left\|\sin\Te (\hbV,\bV)\right\| \leq C \lkv \frac{K^{3/2}}{\sqrt{n \, \rhon\, L }} + \frac{K^{5/2}}{n \, \rhon} \rkv,
\ee 
while \cite{MinhTang_arxiv2022}, who use a different technique for   recovery of $\bV$, 
state that, with high probability,    
$\|\sin\Te (\hbV,\bV)\|_{2 \to \infty} \leq C\, K n^{-1}\,\sqrt{\log n} / \sqrt{\rhon}$.
The latter leads to 
$\|\sin\Te (\hbV,\bV)\|_F \leq C\, K  \,\sqrt{(n \rhon)^{-1}\, \log n}$. 
Thus, the  upper bound \eqref{eq:M1_COSIE}   is similar to our upper bound \eqref{eq:M_eq_1}, which is derived for the (larger) Frobenius norm and
holds not only in expectation but with the high probability. The upper bound of  \cite{MinhTang_arxiv2022} is larger (if one uses the   Frobenius  norm)
and, in addition, does not decline when $L$ grows.

As our theory (Theorems~\ref{th:error_between} and \ref{th:error_Vm_est}, and also Corollary~\ref{cor:error_within})
the simulation results imply, when $K$ and $M$ are fixed constants, the clustering precision in
both algorithms cease to decrease for a given number of nodes $n$   when  $L$ grows:
\bes
R_{BL}  \lesssim C  \rhon^{-1} n^{-1}, \quad
R_{S,max} \asymp  R_{S,ave} \asymp  R_{WL}  \lesssim C \lkr \rhon^{-1} n^{-1} +  n^{-1}\,  L^{-1}\,  \rhon^{-1}\, \log n \rkr
\ees  
We believe that this is not caused by the  deficiency of our methodology but is rather due to the fact, that the number of 
parameters in the model  grows linearly in $L$ for a fixed $n$. Indeed, even in the case of the simplest, SBM-based DIMPLE model, 
 the  total number of independent
parameters in the model is   $O(K^2 L + M n \log K + L \log M)$, since we have   $L$ matrices $\boB\upl$, 
$M$ clustering matrices for the SBMs in the groups of layers, and  a clustering matrix of the layers,
while the total number of observations is   $O(n^2 L)$. The latter implies that, 
while for small values of $L$, the term $(M n \log K)/(n^2 L)$ may dominate the error,
eventually, as $L$ grows, the term $L(K^2 + \log M)/(n^2 L)$ becomes larger for a fixed $n$.

Incidentally, we observe that a similar phenomenon holds in the MMLSBM, where the block probability matrices are 
the same in all layers of each of the   groups. While \cite{Stanley_2016} does not produce relevant theoretical results, 
 \cite{TWIST-AOS2079} simply assume that $L \leq n$, which makes the issue of error rates for a growing value of $L$ inconsequential. 
Similarly,     the ALMA clustering error rates in \cite{fan2021alma}
\beqns
R_{BL}^{ALMA} & \lesssim & C \lkr \rhon^{-1}\,  n^{-2} + \rhon^{-2}\, n^{-2}\,  [\min(n,L)]^{-1}\rkr, \\
R_{WL}^{ALMA} & \lesssim & C \lkr n^{-1}\,  L^{-1}\,  \rhon^{-1} + \rhon^{-1}\,  n^{-2} + \rhon^{-2}\, n^{-2}\,  [\min(n,L)]^{-1}\rkr,
\eeqns
imply that, for   given $n$ and $\rhon$, as $L$ grows, the clustering errors flatten.

Our simulation study also exhibit similar dynamics. In particular,    the between-layer clustering errors flatten when $n$ is fixed and $L$ grows,
while the errors of subspace estimation and of the within-layer clustering, for a fixed $n$,  decrease initially and then stop decreasing 
as $L$ become larger and larger.

We remark that, unlike the ALMA methodology in \cite{fan2021alma} or the TWIST algorithm in \cite{TWIST-AOS2079},
 all three algorithms in this paper are not iterative.  
It is known, that if one needs to recover a low rank  tensor,  then the power iterations can improve 
precision guarantees. This has been shown   in the context of estimation of a low rank tensor  
in, e.g.,  \cite{8368145}, and in the context of the clustering in the tensor block model in \cite{han2021exact}.
While both ALMA and TWIST are designed for the MMLSBM, which results in a low rank probability tensor, 
the DIMPLE model does not lead to a low rank probability tensor. Therefore,  it is not clear whether iterative techniques are
advantageous in the DIMPLE setting. Our very limited experimentation with iterative algorithms did not lead to
significant improvement of clustering precision. Investigation of this issue is a matter of future research.




%

 
\section{Appendix: proofs and  additional simulations}
\label{sec:appendix}


\subsection{ Proof of Theorem~\ref{th:error_between}  } 
\label{sec:proof_between}

Use notations of the paper, note that
\bes
\left\| \hbU_{A,l} (\hbU_{A,l})^T -  \bU_{P,l} (\bU_{P,l})^T  \right\|^2_F =
2\, \left\| \sin\bTe(\bU_{P,l}, \hbU_{A,l}) \right\|^2_F
\ees 
where $\hbU_{A,l}$ and $\bU_{P,l}$ are defined in \eqref{eq:svd_A_l} and \eqref{eq:svd1},
respectively. By Davis-Kahan Theorem,
\bes
\left\| \hbU_{A,l} (\hbU_{A,l})^T -  \bU_{P,l} (\bU_{P,l})^T  \right\|_F 
\leq \frac{ 2\, \sqrt{K_m} \left\| \bA\upl - \bP\upl   \right\|}{\sig_{K_m} (\bP\upl)}, \quad m = c(l)
\ees 
By Theorem~5.2 of \cite{lei2015}, if $n \rho_n \geq C_{\rho} \log n$, then, for any $\tau >0$, there exists a constant 
$C_{\tau}$, such that 
\bes 
\PP \lfi  \| \bA\upl - \bP\upl  \|  \leq C_{\tau} \sqrt{n \rho_n} 
\rfi  \geq 1 - n^{-\tau}
\ees 
Then
\bes 
\PP  \lfi \max_{\linL}\ \| \bA\upl - \bP\upl  \|  \leq C_{\tau} \sqrt{n \rho_n}
\rfi  \geq 1 - L n^{-\tau}
\ees 
In order to construct a lower bound for $\sig_{K_m}(\bP\upl)$, note that  
under  Assumptions {\bf A1}--{\bf A6}, one has
\be  \label{eq:sig_min} 
\sig_{K_m}(\bP\upl) = \sig_{K_m}(\bQ\upl) \geq C_\lam \, K_m^{-1/2}\,  \|\bQ\upl\| \geq \lowc_{\rho} C_\lam \, K^{-1/2}\, \rhon\, \|\bP_0\upl\| 
\geq \lowc_{\rho} C_\lam  C_{0,P} \, n\, \rhon\, K^{-1} 
\ee  
Combining the formulas and taking into account that 
\bes
 \|\hbTe - \bTe\|^2_F \leq L \max_{\linL}\ \left\| \hbU_{A,l} (\hbU_{A,l})^T -  \bU_{P,l} (\bU_{P,l})^T\right\|_{F} ^2, 
\ees
obtain
\bes 
\PP \lfi  \left\| \hbTe - \bTe   \right\|^2_{F} \leq 
C\, \frac{L K^3}{ n \rho_n}  \rfi \geq 1 - L n^{-\tau}
\ees 
Also, by Davis-Kahan Theorem,
\bes
\| \sin\bTe (\hbcalW, \bcalW) \|_F  \leq  
\frac{\left\| \hbTe - \bTe   \right\|_{F}}{\sig_{M}(\bTe)}
\ees
By formula \eqref{eq:bTe_1} and \eqref{eq:lembF_1},
\bes
\sig_{M}(\bTe) \geq 
\sqrt{\frac{\lowc}{\highc}}\, \frac{1}{\kappa_0^2}\, \left\|  \bTe  \right\|  \geq
\sqrt{\frac{\lowc}{\highc}}\, \frac{1}{\kappa_0^2}\, \frac{\left\|  \bTe  \right\|_F}{\sqrt{M}} \geq
\sqrt{\frac{\lowc}{\highc}}\, \frac{\sqrt{K L}}{\kappa_0^2 \sqrt{M}}
\ees
Hence,
\bes 
\PP \lfi \left\| \sin\bTe (\hbcalW, \bcalW) \right\|^2_F \leq 
\frac{C K^2 M}{ n \rho_n}  \rfi \geq 1 - L n^{-\tau}
\ees 
Use Lemma C.1 of \cite{lei2021biasadjusted}:

\begin{lem} \label{lem:lei2021} {\bf ( Lemma~C.1 of \cite{lei2021biasadjusted}). }
 Let $\bX$ be an $m \times d$ matrix with $K$ distinct rows and minimum pairwise Euclidean norm separation $\ga$.
Let  $\hbX$ be another $(m \times d)$ matrix  and $(\hbTe, \hbQ)$ be an $(1 + \eps)$-approximate  solution
to $K$-means problem with input $X$. Then, the number of errors in $\hbTe$ as an estimate number of errors in $\hbTe$
as an estimate of row clusters of $X$ is no larger than $\bC_{\eps} \left\| \sin\bTe (\hbX, \bX) \right\|^2_F \ga^{-2} $,
where $\bC_{\eps}$ depends only on $\eps$.
\end{lem}

\noindent
Since the row separation of $\bcalW$ is at least $1/\sqrt{L_m}  \geq  \sqrt{M}/(\highc \sqrt{L})$, 
the number of errors is bounded above by $C K^2 L\, (n \rho_n)^{-1}$,
with probability at least $1 - L n^{-\tau}$. 
The latter, in combination with \eqref{eq:nLtau}, implies \eqref{eq:error_between}.


\subsection{ Proof of Theorem~\ref{th:error_Vm_est}  }   
\label{sec:proof_error_Vm_est}

The proof requires the following lemma.
\\

\begin{lem} \label{lem:hatWerr}  
Let $\bW$ and $\hbW$ be defined as in \eqref{eq:bC_structure}  and \eqref{eq:hatW}, respectively. 
Let assumptions of Theorem~\ref{th:error_Vm_est} hold.
Then, on the set $\Om$, with $\PP(\Om) \geq  1 -  L\,  n^{- \tau}$, on which \eqref{eq:error_between} holds,
one has 
\begin{align} \label{eq:LmhLm} 
& 0.5\, L_m^{-1}  \leq \hL_m \leq 2\, L_m^{-1}, \quad \minM \\
\label{eq:sqrtLm} 
& |\hL_m^{-1/2} - L_m^{-1/2}| \leq C (n \rhon \sqrt{L_m})^{-1}\, M K^2, \quad \minM \\ 
\label{eq:hatW_err}
& \min_{\scrP \in \mathfrak{F}  (M)}\  \|\hbW  - \bW\, \scrP \|^2_F \leq C (n \rhon)^{-1}\, M K^2  
\end{align}
\end{lem}

\medskip


\noindent
Consider tensors $\calG \in \RR^{n \times n \times L}$ and $\calH = \calG \tim \bW^T \in \RR^{n \times n \times M}$ 
with layers, respectively, $\bG\upl = \calG(:,:,l)$ and $\bH\upm = \calH(:,:,m)$  of the forms
\be \label{eq:calG}
\bG\upl =   (\bP\upl)^2, \quad \bH\upm = L_m^{-1/2}\, \sum_{c(l)=m} \, \bG\upl, \quad
\linL,\ \minM
\ee
In order to assess $R_{S,max}$ and $R_{S,ave}$, one needs to examine the spectral structure of matrices  $\bH\upm$ 
and their deviation from the sample-based versions $\hbH \upm = \hcalH (:,:,m)$.
We start with the first task.

It follows from \eqref{eq:expans1} and \eqref{eq:Ql} that   
\be \label{eq:bHm_alt}
\bH\upm = \bV\upm\,  \barbQ\upm\, (\bV\upm)^T \quad \mbox{with} \quad 
\barbQ\upm = L_m^{-1/2}\, \sum_{c(l)=m} \, \lkr \bQ\upl \rkr^2
\ee
Here, by \eqref{eq:expans1}, one has $(\bQ\upl)^2 = \bO_Q\upl (\bS_Q\upl)^2 (\bO_Q\upl)^T$,
so that all eigenvalues of $(\bQ\upl)^2$ are positive.  
Applying the Theorem in Complement 10.1.2 on page 327 of \cite{Rao_Rao_1998}
and Assumptions~{\bf A1}--{\bf A6}, obtain
\begin{align*}
\sig_{K_m} (\bH\upm) & = \sig_{K_m} \lkr \barbQ\upm \rkr \geq  L_m^{-1/2}\, \sum_{c(l)=m} \sig_{K_m} \lkr (\bQ\upl)^2 \rkr \nonumber \\
& \geq C_\lam^2\, L_m^{-1/2}\, K_m^{-1}\, \sum_{c(l)=m} \|\bQ\upl\|^2_F 
\geq C_\lam^2\, L_m^{-1/2}\, K^{-1} \, \sum_{c(l)=m} \rhonl^2\, \|\bP_0\upl\|^2_F \\
& \geq C_\lam^2\, \lowc_\rho^2 \, C_{0,P}^2\, L_m^{-1/2}\, K^{-2} \, n^2\ \rhon^2\, L_m
\end{align*}
so that 
\be \label{eq:sigminHm}
\sig_{K_m} \lkr \bH\upm \rkr  \geq C\, (K^2 \sqrt{M})^{-1}\, n^2\ \rhon^2\, \sqrt{L}
\ee
 Using Davis-Kahan theorem, Lemma 1 of \cite{cai2018} and formula  \eqref{eq:sigminHm}, obtain 
\be \label{eq:sinTe_errbound}
\left\| \sin\Te \lkr \hbV\upm, \bV\upm \rkr \right\|_F \leq 
\frac{2 \, \sqrt{2 K_m}\,  \|\hbH\upm - \bH\upm\|}{\sig_{K_m} (\bH\upm)}
\leq   \frac{C\, K^{5/2}\, M^{1/2}\,  \|\hbH\upm - \bH\upm\|}{n^2\, \rhon^2\,  \sqrt{L}}   
%
\ee

Recall that $\bH\upm = [\calG \tim \bW^T](:,:,m)$ and $\hbH\upm = [\hcalG \tim \hbW^T] (:,:,m)$.
Denote 
\be   \label{eq:barbG}
\barbG\upm = \displaystyle \sum_{c(l) =m}  \bG\upl = \sqrt{L_m}\, \bH\upm, \quad 
\hbarbG\upm =  \sqrt{L_m}\, \lkv \hcalG \tim \bW^T \rkv (:,:,m) = \displaystyle \sum_{c(l) =m}  \hbG\upl
\ee  
Observe that  
\be   \label{eq:hbHupm_bHupm}   
  \|\hbH\upm - \bH\upm\|     \leq    \Del_{1}\upm + \Del_{2}\upm,  \quad  
\summM \|\hbH\upm - \bH\upm\|^2 \leq  2( \Del_1 + \Del_2)
\ee
where
\begin{align*}
\Del_{1}\upm & = L_m^{-1/2} \left\|\hbarbG\upm - \barbG\upm \right\|, \quad 
 \Del_{2}\upm   =  \left\| [\hcalG \tim (\hbW  -\bW)^T] (:,:,m)  \right\|, \\
  \Del_i & = \sum (\Del_{i}\upm )^2, \quad i=1,2
\end{align*}
%
To upper-bound $\Del_1\upm$ and $\Del_2\upm$, we use   the following lemma that modifies upper bounds in 
Theorem~3 of  \cite{lei2021biasadjusted} in the absence of  the sparsity assumption $\rhon n \leq C$:  
\\

\begin{lem} \label{lem:lei_modified}  
Let   Assumptions {\bf A1}--{\bf A6} hold,  $\bG\upl = (\bP\upl)^2$ and 
$\hbG\upl = (\bA\upl)^2 - \diag(\bA\upl \bone)$, where $c(l) = m$, $l \in [\tilL]$.
Let 
\bes 
\bG = \sum_{l=1}^{\tilL}\, \bG\upl, \quad  \hbG = \sum_{l=1}^{\tilL}\,  \hbG\upl 
\ees
Then, for any $\tau >0$, there exists a constant $C$  that depends  only on   constants  
in Assumptions {\bf A1}--{\bf A6}, and a constant $\tilC_{\tau, \eps}$ which depends  only on $\tau$ and 
$\eps$, such that one has 
\be \label{eq:lei_modified}
\PP \lfi \|\hbG - \bG\|  \leq C\, \lkv \rhon^{3/2} n^{3/2} \sqrt{\tilL\, \log (\tilL + n)} + \rhon^2 n  \tilL  \rkv \rfi 
\geq 1 - \tilC_{\tau, \eps} n^{1 - \tau}
\ee
\end{lem}

\medskip

\noindent
Applying Lemma~\ref{lem:lei_modified} with $\tilL = L_m$ and taking into account that, 
by assumption \eqref{eq:nLtau}, one has $\log n \leq \log(L+n) \leq (1 + \tau_0) \log n$,
obtain that, with probability at least 
$1 - \tilC_{\tau, \eps} n^{1- \tau}$, one has
$\Del_{1}\upm \leq \tilC  [\rhon^{3/2} n^{3/2}  \sqrt{\log n} + \rhon^2\, n  \sqrt{L_m}]$.
Therefore, 
\be \label{eq:Del1_upm}
\PP \lfi \max_{\minM} \Del_1\upm  \leq C \lkv \rhon^{3/2} n^{3/2}  \sqrt{\log n} + \rhon^2\, n  \sqrt{L/M}\rkv \rfi 
\geq 1 - \tilC_{\tau, \eps}\, M\, n^{1 - \tau}
\ee
and $\Del_1 \leq C [\rhon^{3/2} n^{3/2} \, M \sqrt{\log n} + \rhon^2\, n  \sqrt{L\, M}]$ with the same probability.

In the case of $\Del_{2}\upm$, we start with an upper bound for $\Del_2$. 
Note that, by Cauchy inequality and Lemma~\ref{lem:hatWerr}, with probability at least $1 -  L\,  n^{- \tau}$, one has
\beqn  
\Del_2 & = & \summM \left\|\sumlL \hbG\upl (\hbW_{l,m} - \bW_{l,m}) \right\|^2 \leq 
\summM \lkv \sumlL \left\| \hbG\upl\right\| \, \left|\hbW_{l,m} - \bW_{l,m}\right| \rkv^2 \nonumber \\
\label{eq:Del2_v1}
& \leq  & 
\|\hbW  - \bW\|_F^2 \ \sumlL \| \hbG\upl\|^2   \leq 
C (n \rhon)^{-1}\, M\, K^2\ \sumlL \| \hbG\upl\|^2 
\eeqn 
In order to obtain an upper bound for the sum of $\| \hbG\upl\|^2$, use Lemma~\ref{lem:lei_modified} with $\tilL = 1$. 
Derive 
\bes  
\PP \lfi \max_{\linL}\ \|\hbG\upl - \bG\upl \|^2  \leq C\, \lkv \rhon^3 n^3  \, \log n + \rhon^4 n^2    \rkv \rfi 
\geq 1 - \tilC_{\tau, \eps} L\, n^{1 - \tau}
\ees
On the other hand,  
\beqns
\|\bG\upl\| \leq \|\bP\upl\|^2 \leq  C_\lam^{-2}\, [\sig_{K_m}(\bP\upl)]^2 \leq  C_\lam^{-2}\, K_m^{-1} \|\bP\upl\|^2_F
\leq  C_\lam^{-2}\, C_K^{-1}\, K^{-1} (n \rho_n)^2
\eeqns 
Since $\|\hbG\upl\| \leq  \|\bG\upl\| + \|\hbG\upl - \bG\upl \|$,   with probability at least 
$1 - \tilC_{\tau, \eps} L\, n^{1 - \tau}$, obtain
\bes
\max_{\linL}\  \|\hbG\upl\|^2 \leq C \lkr K^{-2} \, n^4\,  \rho_n^4 + \rhon^3 n^3  \, \log n + \rhon^4 n^2 \rkr
\leq C K^{-2} \, n^4\,  \rho_n^4 \lkr 1 + (n \rhon)^{-1} \, K^2 \log n \rkr
\ees
Plugging the latter upper bound into \eqref{eq:Del2_v1}, obtain 
\be \label{eq:Del2}
\PP \lfi \Del_2 \leq C n^3 \rhon^3 \, L\, M\, \lkr 1 +  (n \rhon)^{-1} \, K^2 \log n \rkr \rfi
\geq 1 - \tilC_{\tau, \eps}\, L\, n^{1 - \tau}
\ee
To complete the proof, combine formulas  \eqref{eq:sinTe_errbound}, \eqref{eq:hbHupm_bHupm},
\eqref{eq:Del1_upm} and  \eqref{eq:Del2} take into account that $\Del_2\upm \leq \sqrt{\Del_2}$
for any $\minM$.


\subsection{ Proof of Corollary~\ref{cor:error_within}  }
\label{sec:proof_within_alt}

To find the clustering errors for each group of clusters, we again use Lemma~\ref{lem:lei2021}
which yields that the number of clustering errors in the layer $\minM$ is bounded above by 
 $\bC_{\eps} \left\| \sin\bTe (\hbV\upm, \bV\upm) \right\|^2_F \ga_m^{-2}$, where $\ga_m$
is the minimum pairwise Euclidean norm separation between rows of matrix $\bV\upm$. 
It is easy to see that under Assumptions {\bf A1}--{\bf A6}, one has
\be \label{eq:gammam}
\gamma_m^2 \geq 2 \min(n_{k,m}^{-1}) \geq  2\, C_K\, K/(\highc \, n), 
\ee
so that  the total number of errors is bounded above by $C \, M \, K^{-1}\, n\, R_{S,ave}$
where  $R_{S,ave}$ is given by \eqref{eq:error_Save}. Then, the average within layer clustering error
is bounded above by $ K^{-1}\,  R_{S,ave}$, which completes the proof.   



\subsection{ Proof of supplementary lemmas  }
\label{sec:proof_lemmas}


{\bf Proof of Lemma~\ref{lem:bF_struct} }  
Note that, due to the structure of the tensor $\calB$, for some $s>0$, one has 
$\sig_{\min} (\barbR) = \sig_{\max} (\barbR) = s$, so that 
\bes
\sig_1 (\bF) \leq \sig_1^2 (\barbD) \, s\, \sqrt{\max_{\minM} L_m}, \quad
\sig_M (\bF) \geq \sig_M^2 (\barbD) \, s\, \sqrt{\min_{\minM} L_m}. 
\ees
Then, by Assumptions {\bf A1}  and {\bf A4}, 
$\sig_1^2(\bF) \leq \kappa_0^4 \sig_M^2(\bF) \highc/\lowc$. 
Therefore, the first inequality in \eqref{eq:lembF_1} holds.
To prove the second inequality, observe that 
\bes  
\|\bTe\|^2_F = \Tr(\bF \bF^T (\barbU^T \barbU \otimes \barbU^T \barbU)) = \|\bF\|^2_F
\ees 
and, on the other hand,
\be \label{eq:norm_bTe}
\|\bTe\|^2_F =  \sum_{l=1}^L \|\bU_{P,l}  (\bU_{P,l})^T\|^2_F = 
\sum_{m=1}^M   L_m \|\bV\upm (\bV\upm)^T \|^2_F =  \sum_{m=1}^M   L_m K_m \geq C_K K L, 
\ee 
which together complete the proof.
\\

\medskip 


{\bf Proof of Lemma~\ref{lem:hatWerr}   } 
\ignore{
\begin{lem} \label{lem:hatWerr}  
Let $\bW$ and $\hbW$ be defined as in \eqref{eq:bC_structure}  and \eqref{eq:hatW}, respectively. 
Let assumptions of Theorem~\ref{th:error_Vm_est} hold.
Then, on the set $\Om$, with $\PP(\Om) \geq  1 -  L\,  n^{- \tau}$, on which \eqref{eq:error_between} holds
one has 
\begin{align} \label{eq:LmhLm} 
& L_m({-1}/2 \leq \hL_m \leq 2 L_m({-1}, \quad \minM \\
\label{eq:sqrtLm} 
& |\hL_m^{-1/2} - L_m^{-1/2}| \leq C (n \rhon \sqrt{K})^{-1} M K^2, \quad \minM \\ 
\label{eq:hatW_err}
& \min_{\scrP \in \mathfrak{F}  (M)}\  \|\hbW  - \bW\, \scrP \|^2_F \leq C (n \rhon)^{-1}\, M K^2  
\eeqn
\end{lem}
} 
Note that, for $\minM$, $|\hL_m - L_m| \leq L R_{BL} \leq C (n \rhon)^{-1}\,L\,K^2$.
Then,
\bes   
\left| \frac{1}{\hL_m}  -  \frac{1}{L_m}\right| = \frac{ |\hL_m  - L_m|}{\hL_m\, L_m} 
\leq \frac{C M K^2}{n \rhon}\  \frac{1}{\hL_m}
\ees
Then, due to assumption \eqref{eq:nKMcond}, the coefficient in front of $\hL_m^{-1}$ 
is bounded by $1/2$ and, hence,   \eqref{eq:LmhLm}  holds.
Inequality \eqref{eq:sqrtLm} follows directly from the upper bound on  $|\hL_m - L_m|$ 
and \eqref{eq:LmhLm}.

To prove \eqref{eq:hatW_err}, recall that formulas \eqref{eq:bC_structure} and \eqref{eq:hatW} imply that
\beqn 
\|\hbW - \bW \|^2_F & \leq & \left \|\hbC (\hbD_{\hat{c}})^{-1/2} - \bC (\bD_{c})^{-1/2} \right \|^2_F \label{eq:hbw} \\
& \leq & 2 \left\|\hbC (\hbD_{\hat{c}})^{-1/2}\right \|^2\ \left \|\bI_M - (\hbD_{\hat{c}})^{1/2} (\bD_{c})^{-1/2} \right \|^2_F
+ 2 \left\| \hbC - \bC \right\|^2_F \  \left\|(\bD_{c})^{-1/2} \right\|^2  \nonumber
\eeqn 
where $\bD_c = \diag(L_1, ..., L_M)$ and $\hbD_{\hat{c}} = \diag(\hL_1, ..., \hL_M)$.
It is easy to see that $\|\hbC (\hbD_{\hat{c}})^{-1/2}\|=1$   in \eqref{eq:hbw},  and that, by Assumption~{\bf A1},
$\|(\bD_{c})^{-1/2}\|^2 \leq (\min L_m)^{-1} \leq M/(\lowc\, L)$. Also, $\| \hbC - \bC\|^2_F \leq 2 L \ R_{BL}$, and 
\begin{align*} 
\|\bI_M  & -  (\hbD_{\hat{c}})^{1/2} (\bD_{c})^{-1/2}\|^2_F 
  =  \Tr(\bI_M +  \hbD_{\hat{c}} \bD_{c}^{-1} - 2 (\hbD_{\hat{c}})^{1/2} (\bD_{c})^{-1/2}) \\
& = \sum_{m=1}^M \frac{\lkr \hL_m^{1/2} -  L_m^{1/2}\rkr^2}{L_m}  
\leq \sum_{m=1}^M \frac{|\hL_m  -  L_m|}{L_m} \leq \frac{M}{\lowc\, L}\ \sum_{m=1}^M  |\hL_m  -  L_m |,
\end{align*}
due to Assumption~{\bf A1}, and    $(\sqrt{a} - \sqrt{b})^2 \leq |a-b|$   for  any $a, b >0$.
Since $\sum   |\hL_m  -  L_m |$ is dominated by the number of clustering errors $L\, R_{BL}$, plugging all 
components into \eqref{eq:hbw}, obtain \eqref{eq:hatW_err}.
\\

\medskip


{\bf Proof of Lemma~\ref{lem:lei_modified}    }  
Let  $\bX\upl = \bA\upl - \bP\upl$, $l=1, ..., \tilL$. With some abuse of notations, for any square matrix $\bQ$,
let $\diag(\bQ)$ be the diagonal matrix which diagonal entries are equal to the diagonal entries
of $\bQ$, while for any vector $\bq$, let $\diag(\bq)$ be the diagonal matrix with the vector $\bq$ on the diagonal.
Then, $\hbG - \bG = \bS_1 + \bS_2 + \bS_3$ where 
\begin{align*}
& \bS_1 = \sum_{l=1}^{\tilL}\, (\bP\upl \bX\upl + \bX\upl \bP\upl), \quad
  \bS_2 = \sum_{l=1}^{\tilL}\, \lkv (\bX\upl)^2 - \diag((\bX\upl)^2) \rkv, \\
& \bS_3 = \sum_{l=1}^{\tilL}\, \lkv \diag((\bX\upl)^2) - \diag(\bA\upl \bone) \rkv    
\end{align*}
Therefore, $\|\hbG - \bG\|^2 \leq 3(\|\bS_1\|^2 + \|\bS_2\|^2 + \|\bS_3\|^2)$.

To bound above $\|\bS_1\|^2$, $\|\bS_2\|^2$ and $\|\bS_3\|^2$, apply Theorems~2 and 3 
of  \cite{lei2021biasadjusted} with $v_1 = v_2 = 2 \highcrho \rhon$, $R_1=R_2 = R_2' =1$ 
and $v_2' = 2 \highcrho^2 \rhon^2$. Using Theorems~2 with $m=r=n$ and $t^2 = \tau \highcrho^2 C_{\rho} \rhon^3 n^3 \tilL \log n$, obtain
\bes
\PP \lfi \|\bS_1\|^2 \leq \tilC \rhon^3 n^3 \tilL \log n \rfi \geq 1 - 4 n ^{1 - \tau}
\ees
The first part of Theorem~3 yields that, due to Assumption~{\bf A3}, 
\bes 
\PP \lfi \|\bS_2\|^2 \leq \tilC \rhon^2 n^2 \tilL \log^2 (n+\tilL) \rfi \geq 1 -  C (n+ \tilL) ^{1 - \tau}
\ees
Now, $\|\bS_3 \| \leq \|\bS_3 - \EE (\bS_3)  \| + \displaystyle \max_i | (\EE \bS_3) (i,i)|$,
since $\bS_3$ is a diagonal matrix. Applying second part of Theorem~3 with $\sig_2 =1$ and 
$\sig'_2 = \sqrt{\tilL n}$, obtain 
\bes 
\PP \lfi \|\bS_3 - \EE (\bS_3) \|^2 \leq \tilC \rhon  n  \tilL \log^2 (n+\tilL) \rfi \geq 1 -  C (n+ \tilL) ^{1 - \tau}
\ees
Finally,
\bes
|(\EE \bS_3) (i,i)|  = \left| \sum_{l=1}^{\tilL} \lkv \EE \sum_{j=1}^n [\bX\upl(i,j)]^2 - \sum_{j=1}^n \bP\upl (i,j) \rkv \right|
= \sum_{l=1}^{\tilL}  \sum_{j=1}^n [\bP\upl (i,j)]^2 \leq \rhon^2 n \tilL,
\ees
which completes the proof.



\subsection{The DIMPLE model versus the MMLSBM}
\label{sec:DINMPLE-MMLSBM}

As we have previously  mentioned, in this paper we consider the DIMPLE model, which is a more general   model 
than the MMLSBM. Specifically, the MMLSBM has only $M$ types of layers in the tensor and, therefore,
results in a low rank tensor. On the other hand, all tensor layers in the DIMPLE model can be different and, therefore,
the tensor is not of low rank. In this section, we carry out a limited simulation study, the purpose 
of which is to convince a reader that, while our algorithms work in the case of the MMLSBM, 
the algorithms designed for the MMLSBM produce poor results when data are generated according to
the DIMPLE models.

In particular, in both scenarios,  we first fix $n$, $L$, $M$, $K$ and generate $M$ 
groups of layers using the multinomial distribution with equal probabilities $1/M$. Similarly,
we generate $K$ communities in each of the groups of layers   using the multinomial distribution 
with equal probabilities $1/K$. In this manner, we obtain  community assignment matrices $\bZ{\upm}$,
$m=1, ...,M$, in each layer $l$ with $c(l)=m$, where $c: [L] \to [M]$ is the layer assignment function. 
Next, we choose sparsity parameters $c$ and $d$ and assortativity parameter $w$.

\begin{figure}[t]  
\label{figa1}
\hspace*{-1.6cm}  
\includegraphics[width=18.5cm, height=4.3cm]{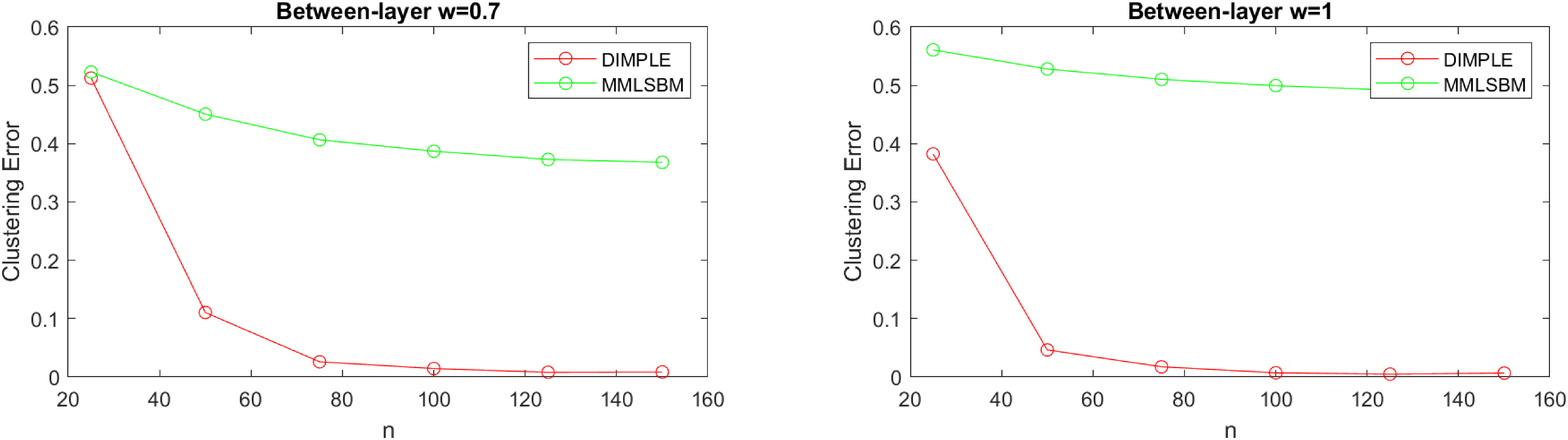}
\hspace*{-1.6cm}  
\includegraphics[width=18.5cm, height=4.3cm]{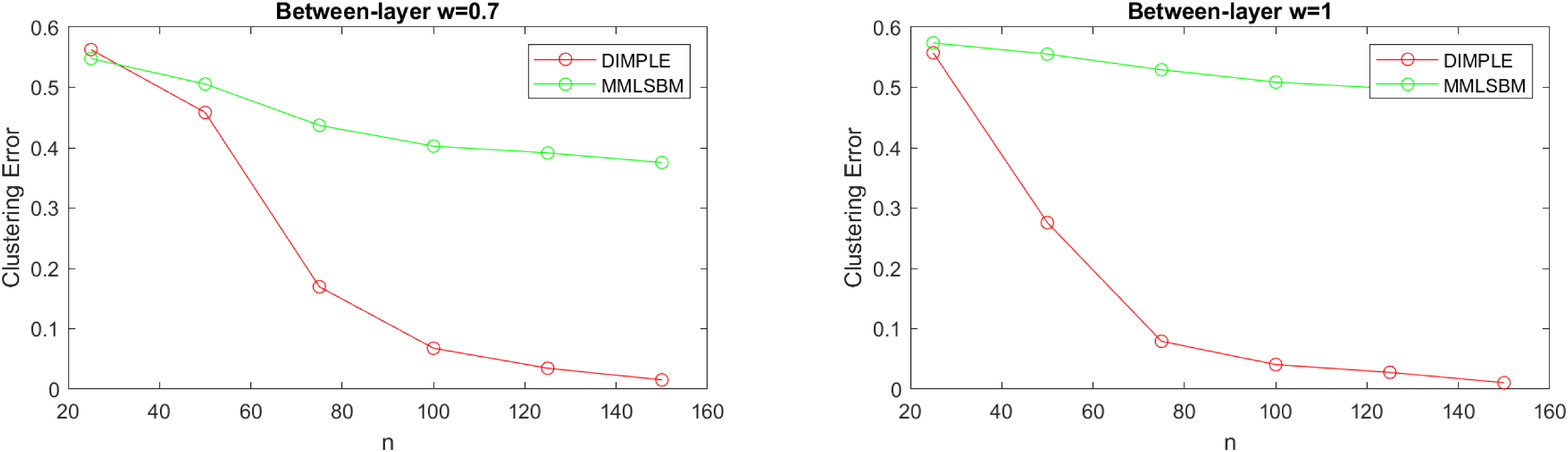}
\caption{{\small The  between-layer clustering error rates of Algorithm 1  and Alternative Minimization Algorithm of \cite{fan2021alma}.
Data are generated using DIMPLE model with $L = 50$, $c = 0$, $d = 0.8$ (top) and  $c = 0$, $d = 0.5$ (bottom),  
and  $w=0.7$ (left panel) or  $w=1$ (right panel).
}}
\end{figure}

\begin{figure}[t]  
\label{figa2}
\hspace*{-1.6cm}  
\includegraphics[width=18.5cm, height=4.3cm]{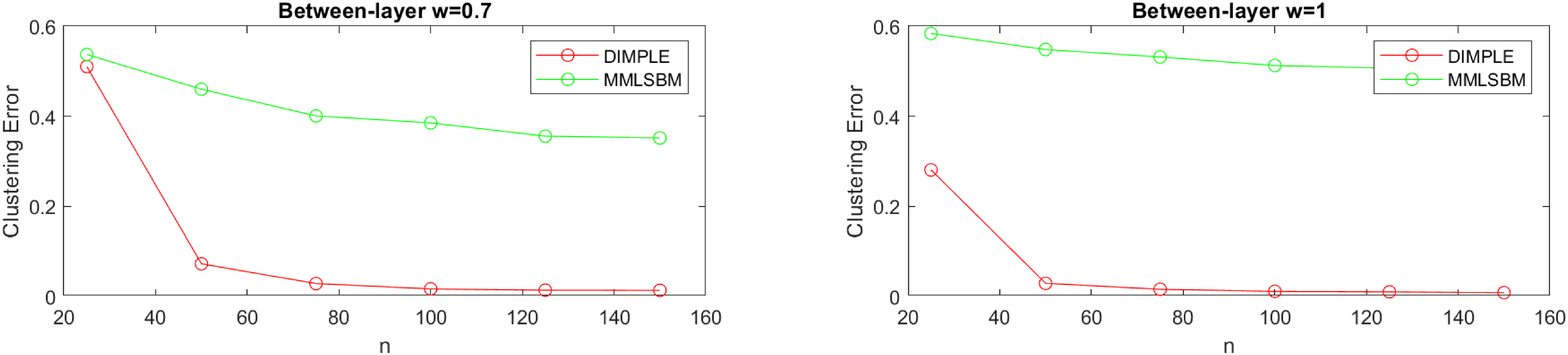}
\hspace*{-1.6cm}  
\includegraphics[width=18.5cm, height=4.3cm]{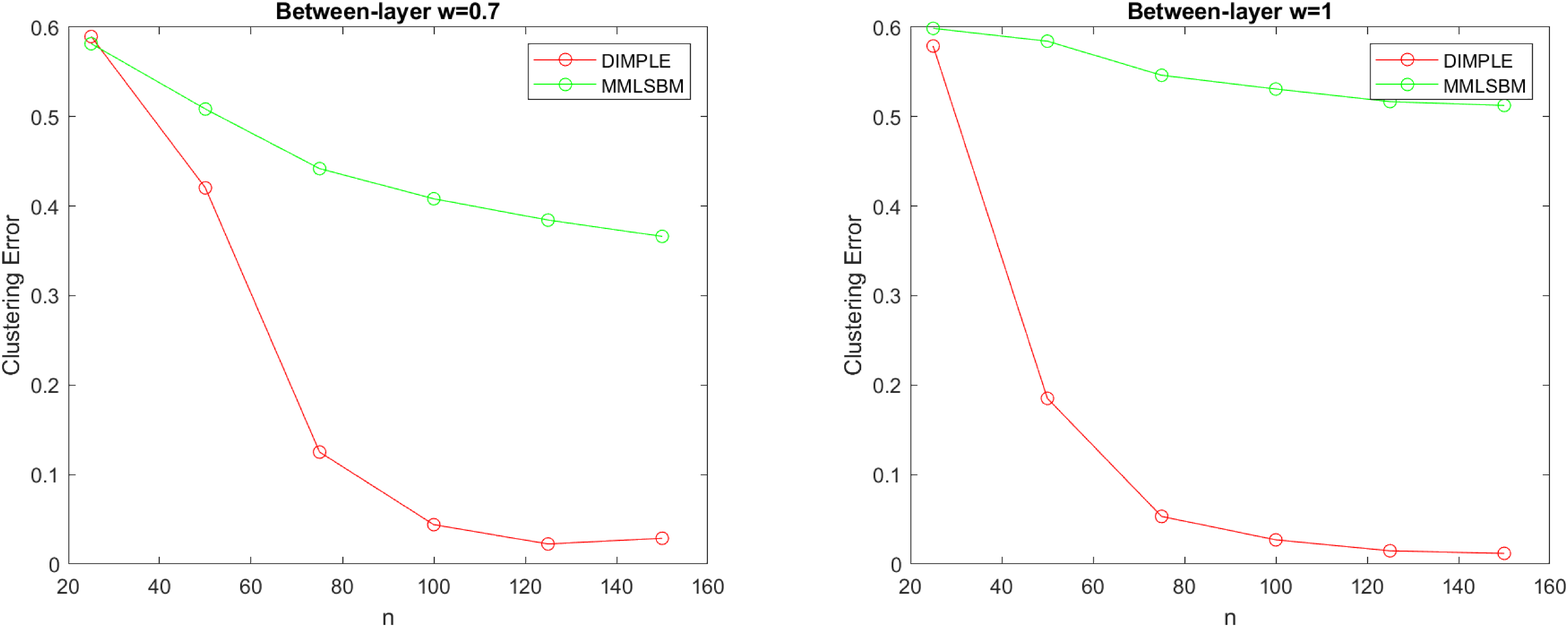}
\caption{{\small The  between-layer clustering error rates of Algorithm 1  and Alternative Minimization Algorithm of \cite{fan2021alma}.
Data are generated using DIMPLE model with $L = 100$, $c = 0$, $d = 0.8$ (top) and  $c = 0$, $d = 0.5$ (bottom), 
$n = 20, 40, 60, 80, 100, 120, 140, 160$ 
and  $w=0.7$ (left panel) or  $w=1$ (right panel).
}}
\end{figure}

In order to generate data according to the DIMPLE model, we obtain the entries of $\bB\upl$, $l=1, ...,L$, 
as uniform random numbers  between $c$ and $d$, and then multiply all the non-diagonal entries of those matrices by $w$.
Therefore, if  $w<1$ is small, then the network is strongly assortative, i.e., there is higher  
probability for nodes in the same community to connect.

The next four figures present simulation results for $K=5$, $M=3$ and various values of $L$, $n$, $c$, $d$ and $w$.
We present only the between layer clustering errors since, in the presence of the assortativity assumption, the within-layer clustering
in the MMLSBM and the DIMPLE model can be carried out in a similar way. We compare the performances of Algorithm~\ref{alg:between}
in this paper with  the Alternative Minimization Algorithm (ALMA) of \cite{fan2021alma}. 

As our simulations  show, when data are generated according to the DIMPLE model, 
Algorithm~1 in our paper allows to reliably separate layers of the network into $M$ types, 
while ALMA fails to do so. The reason for this is that ALMA   expects the matrices of probabilities to be identical in those layers,
although, in reality, they are not. As a result, when $n$ grows, the clustering errors do  not tend to zero but just flatten.

\begin{figure}[t]  
\label{figa3}
\hspace*{-1.6cm}  
\includegraphics[width=18.5cm, height=4.3cm]{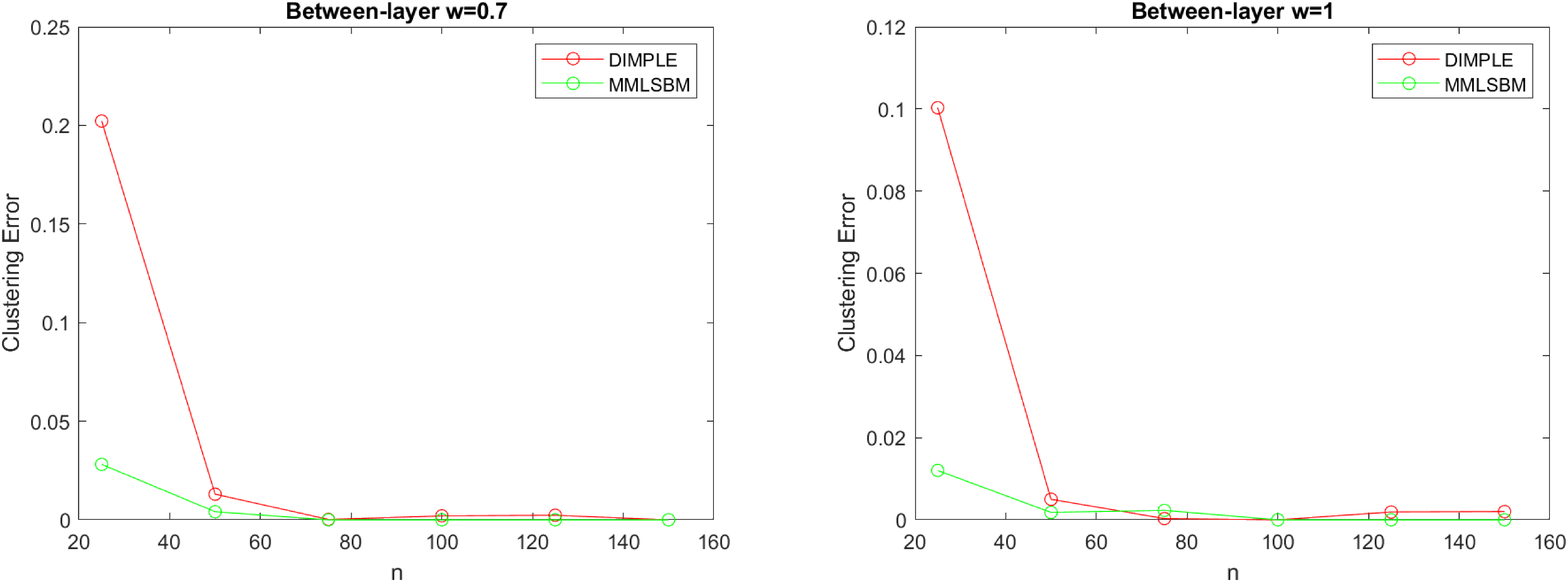}
\hspace*{-1.6cm}  
\includegraphics[width=18.5cm, height=4.3cm]{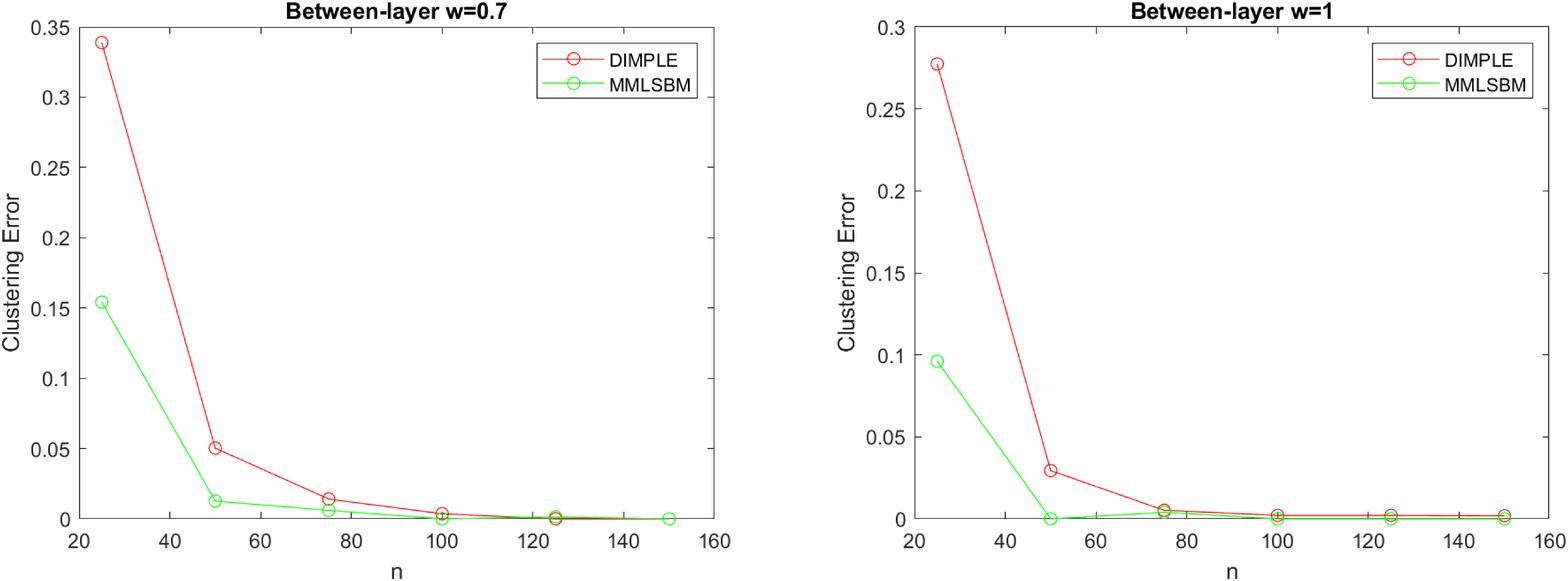}
\caption{{\small The  between-layer clustering error rates of Algorithm 1  and Alternative Minimization Algorithm of \cite{fan2021alma}.
Data are generated using MMLSBM   with $L = 50$, $c = 0$, $d = 0.8$ (top) and  $c = 0$, $d = 0.5$ (bottom), 
$n = 20, 40, 60, 80, 100, 120, 140, 160$ 
and  $w=0.7$ (left panel) or  $w=1$ (right panel).
}}
\end{figure}

\begin{figure}[t]  
\label{figa4}
\hspace*{-1.6cm}  
\includegraphics[width=18.5cm, height=4.3cm]{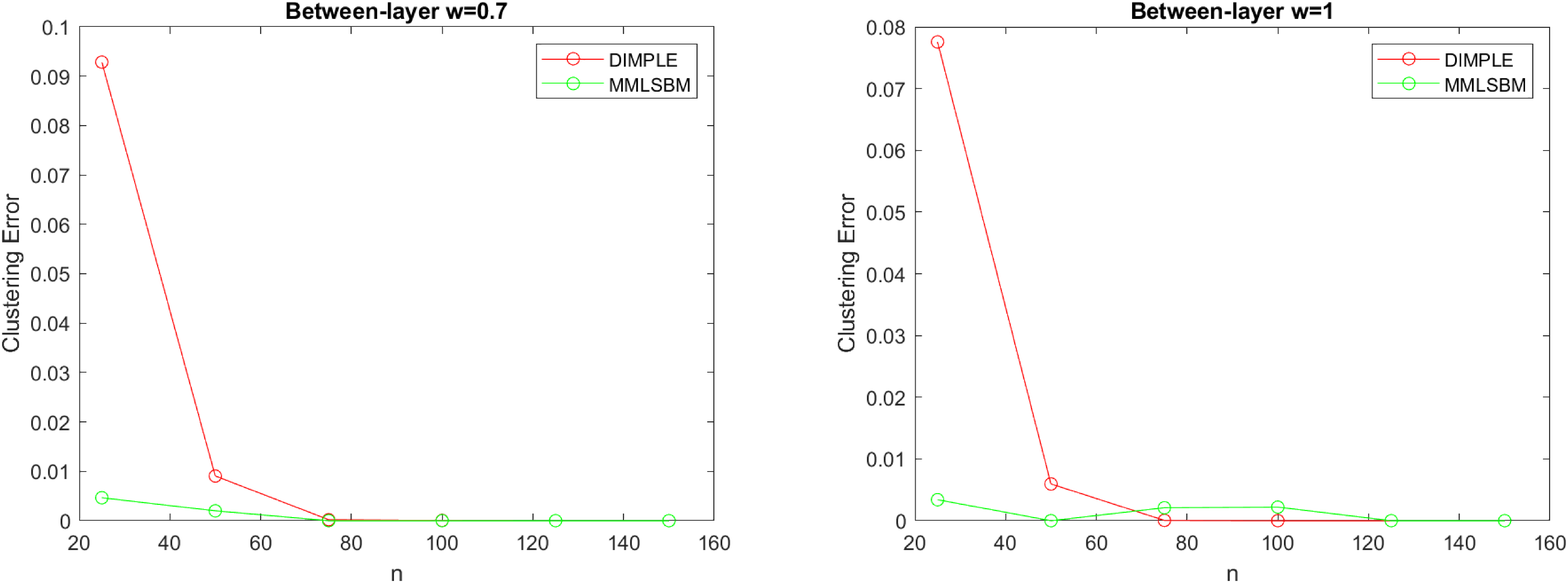}
\hspace*{-1.6cm}  
\includegraphics[width=18.5cm, height=4.3cm]{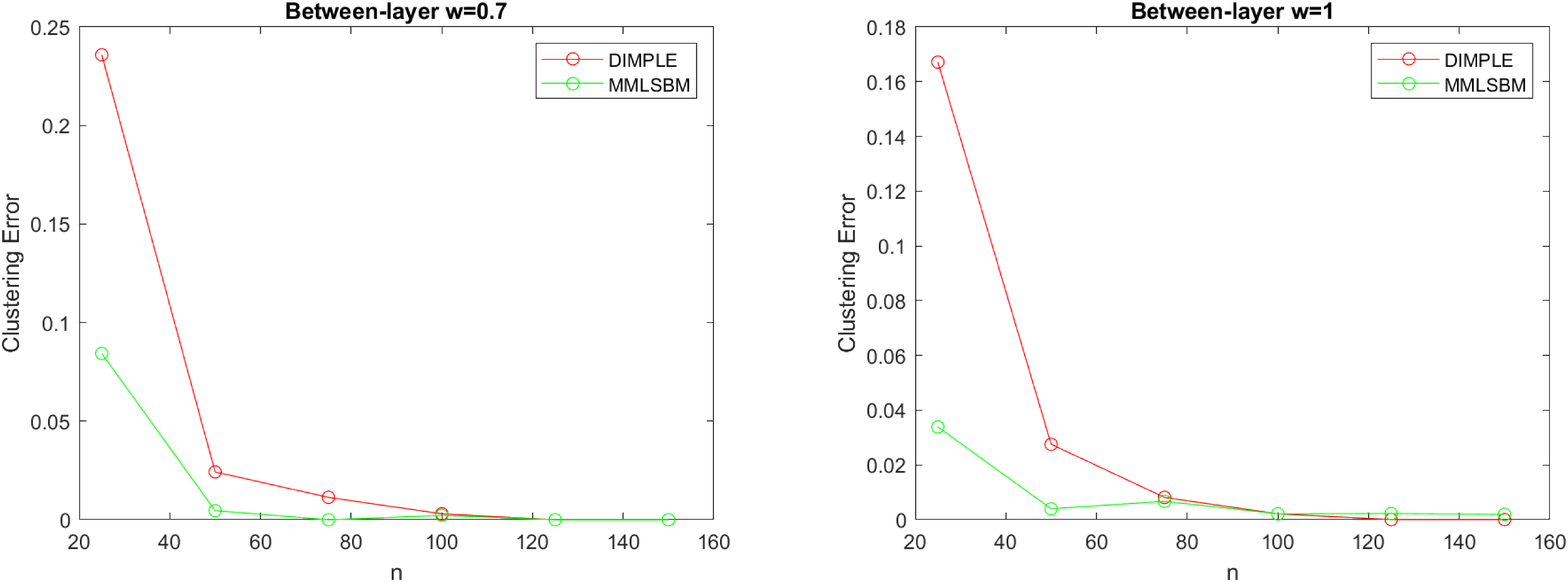}
\caption{{\small The  between-layer clustering error rates of Algorithm 1  and Alternative Minimization Algorithm of \cite{fan2021alma}.
Data are generated using MMLSBM   with $L = 100$, $c = 0$, $d = 0.8$ (top) and  $c = 0$, $d = 0.5$ (bottom),
and  $w=0.7$ (left panel) or  $w=1$ (right panel).
}}
\end{figure}

Next, we generate data according to the MMLSBM. Note that the main difference between the MMLSBM and the DIMPLE model is that in MMLSBM 
one has only $M$ distinct matrices $\bB\upl$, since $\bB\upl = \bB^{(c(l))}$, $l=1, ...,L$. So, in order to generate MMLSBM,
we generate $M$ matrices $\bB\upm$, $m=1, ..., M$, and then set $\bB\upl = \bB^{(c(l))}$, $l=1, ...,L$. 
Figures~\ref{figa1}--\ref{figa4} exhibit results of application of   Algorithm~1  and ALMA of \cite{fan2021alma}
to the generated data sets. As it is expected,  for small values of $n$, ALMA of \cite{fan2021alma} leads to a  better
clustering precision. The latter is due to the fact that Algorithm~1 relies on the SVDs of the 
layers of the adjacency tensor $\calA$, that are not reliable for small values of $n$.
In addition, Algorithm~1 cannot take into account that the probability tensor is of a low rank since this is not true for the DIMPLE model. 
However, these advantages become less and less significant as $n$ grows. As Figures~\ref{figa1}--\ref{figa4}  show, both algorithms have 
similar clustering precision for larger values of $n$, specifically, for $n \geq n_0 $, 
where $n_0$ is between 60 and 100,  depending on a particular simulations setting.

\section*{Acknowledgments} 

Both authors of the paper were  partially supported by National Science Foundation
(NSF)  grant DMS-2014928.



\bibliography{MultilayerNew}

\end{document}